\relax
\documentclass[letterpaper]{article} 
\usepackage{aaai22}  
\usepackage{times}  
\usepackage{helvet}  
\usepackage{courier}  
\usepackage[hyphens]{url}  
\usepackage{graphicx} 
\usepackage{natbib}  
\usepackage{caption} 
\DeclareCaptionStyle{ruled}{labelfont=normalfont,labelsep=colon,strut=off} 
\usepackage{algorithm}
\usepackage{algorithmic}

\usepackage{graphicx}
\usepackage[labelformat = parens, skip=4pt]{subcaption}
\usepackage{multirow}

\usepackage{wrapfig}
\usepackage{booktabs}

\usepackage{amsmath}
\usepackage{amssymb}

\usepackage{xcolor}
\newcounter{HWNumberOfComments}
\stepcounter{HWNumberOfComments}

\newcounter{JSNumberOfComments}
\stepcounter{JSNumberOfComments}

\usepackage{newfloat}
\usepackage{listings}
\floatstyle{ruled}
\newfloat{listing}{tb}{lst}{}
\floatname{listing}{Listing}
\title{Public Perception of Generative AI on Twitter:\\An Empirical Study Based on Occupation and Usage}
\author{
    Kunihiro Miyazaki\textsuperscript{\rm 1},
    Taichi Murayama\textsuperscript{\rm 2}, 
    Takayuki Uchiba\textsuperscript{\rm 3}, 
    Jisun An\textsuperscript{\rm 1}, 
    Haewoon Kwak\textsuperscript{\rm 1}
}
\affiliations{
    \textsuperscript{\rm 1}Indiana University Bloomington,
    \textsuperscript{\rm 2}SANKEN, Osaka University,
    \textsuperscript{\rm 3}Sugakubunka\\

    \{kunihirom,jisun.an,haewoon\}@acm.org, taichi@sanken.osaka-u.ac.jp, takayuki.uchiba@sugakubunka.com\\
}

\begin{document}

\maketitle

\begin{abstract}
The emergence of generative AI has sparked substantial discussions, with the potential to have profound impacts on society in all aspects.
As emerging technologies continue to advance, it is imperative to facilitate their proper integration into society, managing expectations and fear.
This paper investigates users' perceptions of generative AI using 3M posts on Twitter from January 2019 to March 2023, especially focusing on their occupation and usage. 
We find that people across various occupations, not just IT-related ones, show a strong interest in generative AI. 
The sentiment toward generative AI is generally positive, and remarkably, their sentiments are positively correlated with their exposure to AI. 
Among occupations, illustrators show exceptionally negative sentiment mainly due to concerns about the unethical usage of artworks in constructing AI.
People use ChatGPT in diverse ways, and notably the casual usage in which they ``play with'' ChatGPT tends to associate with positive sentiments.
After the release of ChatGPT, people's interest in AI in general has increased dramatically; however, the topic with the most significant increase and positive sentiment is related to crypto, indicating the hype-worthy characteristics of generative AI.
These findings would offer valuable lessons for policymaking on the emergence of new technology and also empirical insights for the considerations of future human-AI symbiosis.
\end{abstract}

\section{Introduction}
Generative AI has garnered significant attention in recent years, demonstrating innovative advancements in various fields, including text, image, and software coding. 
Especially, ChatGPT, a specialized conversational application of the GPT series has gained widespread popularity for its intelligence and seamless conversation capabilities~\cite{WhatisCh30:online}. 
The global debut of ChatGPT has led to substantial societal reactions, such as the official ban in public schools, a global call for a moratorium on developing GPTs, and the national debates about banning access to ChatGPT~\cite{rudolph6war}. \looseness=-1

Generally, it is essential for society to accept emerging technologies properly. 
Excessive expectations for innovative technologies can result in subsequent disappointment and hinder research progress, while unwarranted fear may impede the adoption of beneficial systems~\cite{cave2019scary}. 
Consequently, early understanding of public perception toward new technologies is a priority~\cite{binder2012measuring}.  
Some AI researchers criticized ChatGPT, arguing that people's expectations are driven by hype~\cite{chris2023What}, whereas the mainstream media has perpetuated a narrative suggesting that people are afraid of AI taking their jobs~\cite{PeopleFe45:online}. 
Despite the debate and importance of the topic, there is a lack of quantitative analysis examining people's perceptions of generative AI. \looseness=-1

In this work, we investigate the public perception of generative AI on Twitter, especially focusing on the user's occupation and usage of this technology.
The impact of emerging technologies on various occupations has long been debated especially in the context of job displacement~\cite{eloundou2023gpts}.  
Since the exceptional usability of generative AI extends its reach to non-IT lay people~\cite{WhyaConv90:online} and provides them with numerous firsthand experiences of AI, understanding their perception and reaction toward generative AI would be particularly insightful for practitioners and policymakers considering the future human-AI symbiosis. \looseness=-1

We set the research questions (RQs) below for analysis.

\noindent \textbf{RQ1: Which occupation mention generative AI?}
We examine how users of various occupations get interested in generative AI by comparing the frequency of the tweets about generative AI and randomly sampled tweets.

\noindent \textbf{RQ2: What are the sentiments of different occupations toward generative AI?}
We analyze whether social media users are positive or negative toward generative AI. 
Additionally, we investigate the relationship between the sentiments by occupations and their ``AI exposure score'' from existing research~\cite{webb2019impact,felten2021occupational,felten2023occupational}.

\noindent \textbf{RQ3: How do people interact with generative AI?}
We analyze the types of users' prompts to ChatGPT by examining the images of ChatGPT screenshots they post.

\noindent \textbf{RQ4: Has the emergence of generative AI changed people's perception of AI?}
We conduct a time-series analysis to examine the impact of generative AI's emergence on people's perceptions of AI in general, focusing on changes in tweet volume and sentiment over time.


Our contributions are as follows: (1) To the best of our knowledge, this is the first study to analyze social media perceptions of generative AIs with respect to occupation and use, which would provide policymakers with valuable insights about human-AI relationships.
(2) We propose methodologies for extracting occupations from Twitter user profiles and analyzing texts extracted from ChatGPT images.
(3) We create a comprehensive Twitter dataset focusing on generative AI, which will be publicly available upon publication. \looseness=-1

\section{Background and Related Works}
\subsection{Public Perceptions Toward AI}
Existing studies have conducted surveys about people's perceptions of AI in general~\cite{,HowAmeri78:online,cave2019scary}, 
and~\citet{FromChat60:online} also conducted an early interview survey on generative AI.
Their findings are basically that people are positive about AI development, although they have heard about the risk of job displacement. 
In addition to the survey-based research, \citet{fast2017long} worked on text-based analysis over 30 years of the New York Times data and showed similar insights that the debate about AI has been consistently optimistic, although the concerns about the negative impact of AI on jobs have grown in recent years. \looseness=-1

This study conducts social media analysis focusing on users' perceptions of generative AI, which remains limited in the literature. 
One of the benefits of social media analysis is the ability to analyze sentiment at a detailed topic level.
In addition, people are showcasing their interactions with generative AI on a daily basis~\cite{TheBrill64:online}, enabling in-depth analysis of their engagement with this technology.

\subsection{Social Media Perception of Emerging Technologies}
Social media has been used to understand public perceptions of emerging technologies, such as IoT~\cite{bian2016mining}, self-driving car~\cite{kohl2018anticipating}, and solar power technology~\cite{nuortimo2018exploring}.
These studies found that a majority of the users were basically positive or neutral toward these emerging technologies.
On the other hand, they identified some problematic issues by text analysis, such as privacy and security issues on IoT devices. \looseness=-1

There are also social media studies focusing on the perception of AI~\cite{manikonda2018tweeting} or early investigation on generative AI, especially ChatGPT~\cite{haque2022think,leiter2023chatgpt}.
These studies found that positive and joyful sentiments toward AI were dominating social media. \looseness=-1

Our study differs from prior research in several key aspects. First, our analysis is more comprehensive--we compare multiple generative AIs, conduct a systematic occupational analysis, and benchmark our findings against randomly sampled tweets. 
Additionally, our novel exploration of images about generative AI enables a more in-depth understanding of the users' reactions. \looseness=-1

\section{Dataset Building}

\subsection{Selection of Generative AI Tools}\label{selection}
To generalize our research, we consider various types of generative AI. 
We select them considering functionality and serviceability on social media.
As for functionality, we select generative AI that is conversational (Chat), generates images given prompts (Image), complements codes in programming tasks (Code), and serves as a base model for various application tools (Model).
As a result, we select the generative AI listed in Table~\ref{table:generativeAI}.
Since they include both models and services of generative AI, we use the notation of ``generative AI tools'' for them.

\begin{table*}[!htb]
\centering
\scalebox{0.85}{
\begin{tabular}{lllp{8cm}rr}
\toprule
Category                    & Name             & Release  & Keywords                                                                                                                   & Num (Orig.) & Num (Fin.) \\ 
\hline
\multirow{3}{*}{Chat}       & ChatGPT          & 11/30/2022 & chatgpt, chat gpt                                                                                                          & 2,106,256   & 1,751,065  \\
                            & Bing Chat        & 02/22/2023 & bing chat, bingchat                                                                                                        & 20,867      & 12,285     \\
                            & Perplexity AI    & 12/07/2022 & perplexity ai,   perplexityai, perplexity.ai, perplexityask, perplexity ask                                                & 2,395       & (2,112)    \\ 
\hline
\multirow{7}{*}{Image}      & DALL·E 2         & 07/20/2022 & dall-e-2, dall-e 2,   dall-e2, dall·e 2, dall·e2, dall·e-2, dalle-2, dalle2, (dalle 2)                                     & 196,714     & 105,270    \\
                            & DALL·E           & 01/05/2021 & dall-e, dall·e,   (dalle)                                                                                                  & 303,470     & 139,103    \\
                            & Stable Diffusion & 08/22/2022 & stable diffusion,   stablediffusion                                                                                        & 453,894     & 230,264    \\
                            & Midjourney       & 03/21/2022 & midjourney, (mid journey)                                                                                                & 517,680     & 350,419    \\
                            & IMAGEN           & 11/02/2022 & imagen                                                                                                                     & 173,620     & (2,805)    \\
                            & Craiyon          & 04/21/2022 & craiyon, dall-e-mini,   dall-e mini, dall-emini, dall·e mini, dall·emini, dall·e-mini, dalle-mini,   dallemini, dalle mini & 66,873      & 34,465     \\
                            & DreamStudio      & 08/20/2022 & dreamstudio                                                                                                                & 10,713      & (3,475)    \\ \hline
Code                        & GitHub Copilot   & 10/29/2021 & copilot, co-pilot                                                                                                          & 325,105     & 112,315    \\ \hline
\multirow{4}{*}{Model} & GPT-4            & 03/14/2023 & gpt-4, gpt4, gpt 4                                                                                                         & 304,719     & 245,174    \\
                            & GPT-3.5          & 11/30/2022 & gpt-3.5, gpt3.5, gpt   3.5                                                                                                 & 11,226      & (5,749)    \\
                            & GPT-3            & 06/11/2020 & gpt-3, gpt3, gpt 3                                                                                                         & 317,852     & 237,190    \\
                            & GPT-2            & 02/14/2019 & gpt-2, gpt2, (gpt 2)                                                                                                       & 39,417      & 30,397     \\ \bottomrule

\end{tabular}
}
\caption{Summary of tweets about generative AI tools. Num (Orig.) indicates the initial tweet count, and Num (Fin.) indicates the noise-reduced tweet count. 
The keywords and numbers with parentheses are omitted from the dataset due to their difficulty in distinguishing from homonyms or the scarcity of their volume.
Note that the release date can vary depending on definitions, but we basically chose the beta release date found on the Web.
}
\label{table:generativeAI}
\end{table*}

\subsection{Generative AI Tweets}\label{Generative AI Tweets}

\noindent \textbf{Tweet retrieval.} 
We use Twitter Academic API to retrieve tweets of generative AI tools from January 1, 2019, to March 26, 2023, covering all their release dates. 
Our search keywords (case-insensitive for search on Twitter) are as exhaustive as possible and include variations (Table~\ref{table:generativeAI}). 
We exclude retweets and non-English tweets.  In total, we retrieve 5,118,476 tweets from 1,475,262 users.
Table~\ref{table:generativeAI} summarizes our dataset.  \looseness=-1

\noindent \textbf{Noise removal.}
To ensure the robustness of our analysis, we perform five types of noise removal on the acquired tweets. 
We remove 1) the tweets in which the name of authors or mentioned users include the name of generative AIs, and the text itself does not; 2) the tweets discussing a generative AI tool prior to its release date so that we gain a clearer understanding of the perception toward it; 3) the keywords that result in low tweet volumes (e.g., Perplexity AI, DreamStudio, and GPT-3.5); 4) the tweets containing homonyms, which are often considered difficulties in analyzing social media posts on certain keywords~\cite{miyazaki2022characterizing}; and 5) the tweets generated by bot-like accounts. 

To identify and remove homonyms, we randomly sample 100 tweets from each keyword and manually annotate them to see if they refer to generative AI tools. 
We find that most of the keywords refer to the corresponding generative AI tools with almost 100\% accuracy. 
However, several keywords contain are highly likely homonyms; thus, we remove them, which are annotated with parentheses in Table~\ref{table:generativeAI} (the accuracies and the detailed reasons for the homonyms are omitted for the reason of space).
Among the keywords with homonyms, we retain GitHub Copilot (i.e., copilot and co-pilot) in our dataset as they do not have the alternative in their keywords, unlike other generative AI tools (note that its homonym indicates the second pilot). 
Furthermore, GitHub Copilot is generally a generative AI tool for engineers/researchers, which enables us to analyze the potential difference in responses from engineers/researchers and lay people.
To ensure GitHub Copilot's inclusion, we build a machine-learning model to classify whether a tweet is truly about GitHub Copilot or not. 
We first annotate additional tweets containing the two keywords (400 tweets in total) and fine-tune a RoBERTa-based model~\cite{barbieri-etal-2020-tweeteval} with these annotations. 
The model achieves an F1 score of 0.94 in 5-fold cross-validation, allowing us to predict and remove noise from GitHub Copilot tweets. 
We also remove the tweets with ``Microsoft 365 Copilot'' because this tool is a different tool from GitHub Copilot and is not publicly available at the time of this study. 
Additionally, we find the mixed use of DALL·E and DALL·E 2 in many tweets; thus, we combine them in our analysis and use the notation of ``DALL·E (2)''.  \looseness=-1

We then remove bot-like accounts. 
We use Botometer~\cite{yang2022botometer}, a widely used tool for bot detection that assigns the bot score (0 to 1 scale) to Twitter accounts. 
Using Botometer for all the 1.4M users is difficult and costly due to the API limit.  
Instead, we extract users with more than five tweets in our dataset (127,547 users) to mitigate the impacts of the active bots.
Using a score of 0.43 as a threshold following~\citet{keller2019social}, a relatively conservative setting among previous literature~\cite{rauchfleisch2020false}, we regard 13,315 (10.4\%) users as bots and remove their 872,893 tweets.
\looseness=-1


In total, we obtain 3,065,972 tweets and 1,082,092 users.  
We denote these generative AI-related tweets as $T_{genAI}$ in the rest of the paper. 
The final number of tweets for each generative AI tool is displayed in Table~\ref{table:generativeAI}.  
Note that when we count the tweets for each generative AI tool, we avoid overlap between substring models (e.g., when counting GPT-3, tweets with GPT-3.5 are not included). \looseness=-1

\subsection{Occupation Extraction}
To conduct an occupational analysis of the reaction to the emergence of generative AIs, 
we infer the major occupations of Twitter users from their profiles, 
inspired by~\cite{zhao2021exploring}. 
Although user profiles on Twitter can present issues with self-disclosure and falsehood, we acknowledge these limitations in our study, as also discussed in~\cite{sloan2015tweets}.  \looseness=-1

\begin{figure}[!htbp]
\centering
\includegraphics[width=0.85\linewidth]{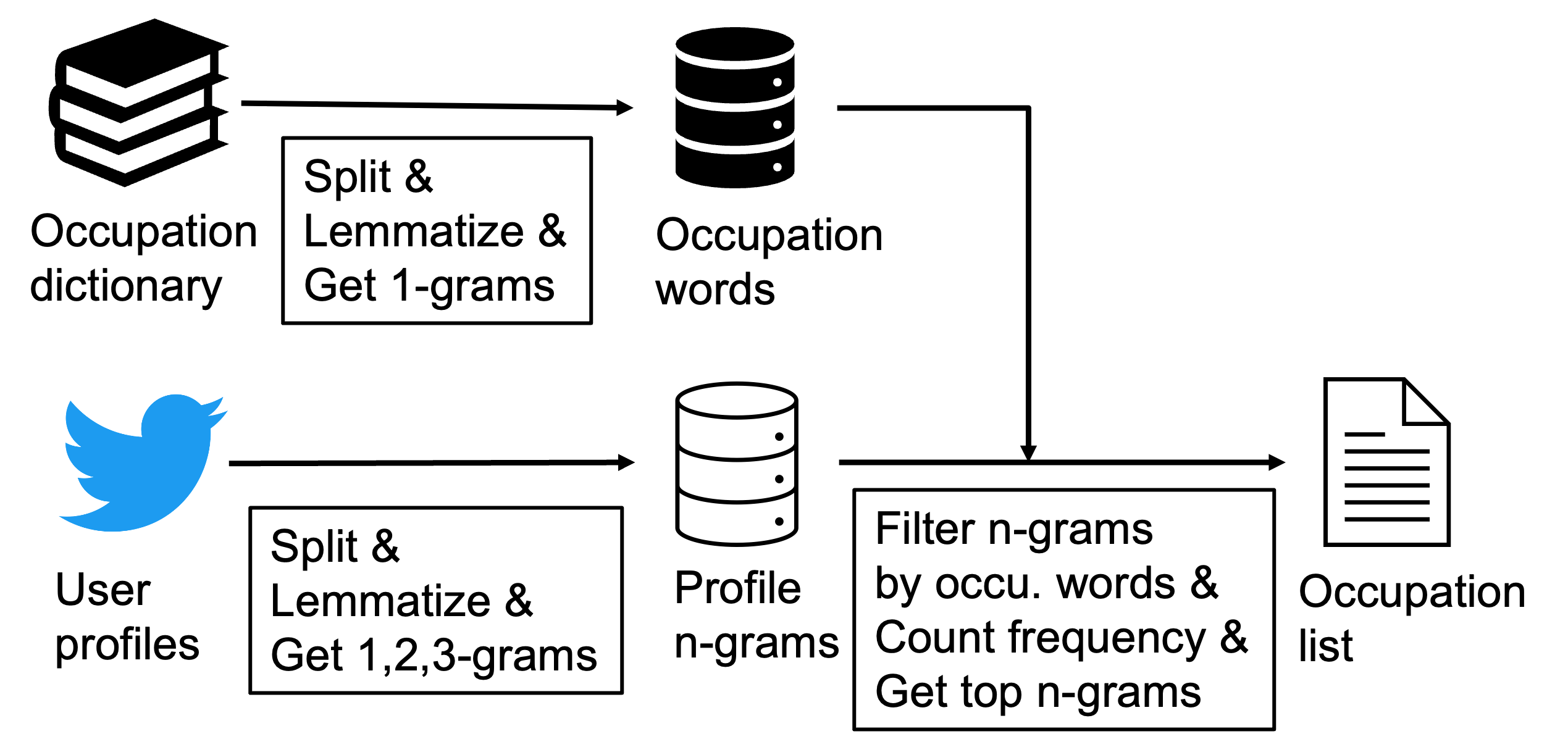}
\caption{
A workflow for extracting major occupations from Twitter user profiles.
}
\label{fig:occupation_extraction}
\end{figure}

We employ keyword-based methods to extract occupations. 
We find the accuracy of machine-learning approaches~\cite{preoctiuc2015analysis,pan2019twitter} is unsatisfactory (about 50\% of accuracy in nine-class classifications).
In contrast, dictionary-based methods offer interpretability and are more commonly used in Twitter research~\cite{sloan2015tweets,hussain2022doctors}. 
Figure~\ref{fig:occupation_extraction} shows our workflow for inferring the occupations of Twitter users.
\looseness=-1

\vspace{1mm}
\noindent \textbf{1. Preparation of dictionaries and user profiles.}
We start with the occupation names from all classes in the Bureau of Labor Statistics (BLS) Occupation Dictionary~\cite{ListofSO56:online}, which is frequently employed for data-driven occupational analysis with O*NET data~\cite{peterson2001understanding}. 
However, it has formal occupational names, failing to capture emerging occupations (e.g., YouTubers). 
To address this limitation, we further use Indeed's occupation list~\cite{FindJobs34:online}, a more up-to-date resource. 
For user data to use for matching, we retrieve the profile texts of all users from $T_{genAI}$. 

\vspace{1mm}
\noindent \textbf{2. Preprocessing for matching occupational names and profiles.}
As the words in the occupation dictionary may not align with the text in the user profiles, and some occupations consist of multiple words, we conduct preprocessing to mitigate their impacts. 
First, we split all occupation names (from the dictionaries) and the profile texts into 1-grams. 
Next, we lemmatize all these 1-grams after removing stop words. 
Finally, we extract unique 1-grams from the dictionary and 1, 2, and 3-grams from the profile texts, designating them as ``Occupation words'' and ``Profile n-grams,'' respectively. 

\vspace{1mm}
\noindent \textbf{3. Extraction of major occupational names.}
To extract the major occupation names in Twitter users' profiles~\cite{zhao2021exploring}, we select the most frequent n-grams related to occupation. 
Initially, we retain only Profile n-grams that contain Occupation words. 
We then count the frequency of the remaining Profile n-grams across all profile texts. 
Then, two authors of this study review the Profile n-grams in descending order of frequency, discussing from the top whether they indicate occupation, and extract the top 30 occupation names, which can be found in Figure~\ref{fig:bar_presence_all}.  \looseness=-1

\subsection{User's Historical Tweets}\label{hist_tweets}
The aims of our RQ4 include measuring shifts in users' attitudes toward AI technology due to the emergence of generative AI. 
To this end, we collect historical tweets of users in $T_{genAI}$.  
We randomly sample 100 accounts from each occupation, ensuring each user belongs to a unique occupation without double belonging.
This process yielded another dataset of 3,387,112 tweets from 3,000 users, including retweets. \looseness=-1

\subsection{Randomly Sampled Tweets}\label{randomsample}
In addition to the tweets collected by Twitter Academic API, we utilize the dataset of 10\% randomly sampled tweets collected by Twitter Decahose API, which spans a three-year period from May 2020 to April 2023, and we detect users' occupations in the same manner as in $T_{genAI}$. 
From this dataset, we extract two subsets.
One is one-tenth of this data (i.e., 1\% sample of Twitter) of all English tweets, which we denote $T_{rand}$ in the rest of the paper. We use $T_{rand}$ as a reference to the usual tweets of occupational users to mitigate the inherent bias of Twitter in our analysis.
The other is the English tweets containing ``\#AI'' and ``\#ArtificialIntelligence,'' where we use all the 10\% sample due to scarcity of the tweets with these hashtags.
We use this AI-related subset as a source to know the perception of AI in general on Twitter (RQ4), considering $T_{genAI}$ does not necessarily reflects the perception of all Twitter users. \looseness=-1

From both of the data, out of the 49,005 accounts that have more than five tweets during the entire period, we remove 15,022 accounts with a Botometer score of 0.43 or higher. 
The volume is 3,601,890 for $T_{rand}$ and 855,589 for AI-related tweets, excluding retweets. \looseness=-1

\subsection{Extraction of Images and Prompts}\label{image_extraction}
\begin{figure}[!ht]
\centering
  \begin{subfigure}[t]{.47\linewidth}
    \centering
    \includegraphics[width=\linewidth]{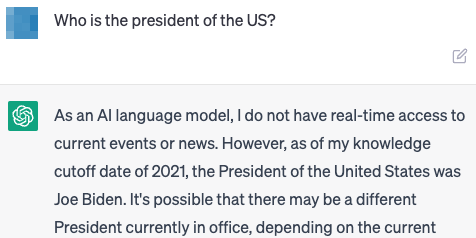}
    \caption{Light mode.}
    \label{fig:1a_sb}
  \end{subfigure}
  \begin{subfigure}[t]{.47\linewidth}
    \centering
    \includegraphics[width=\linewidth]{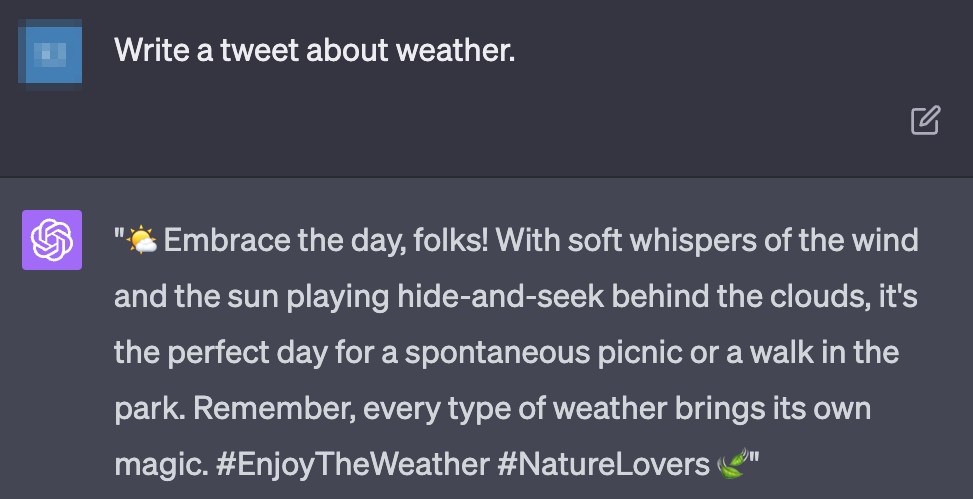}
    \caption{Dark mode.}
    \label{fig:1b_sb}
  \end{subfigure}
  \setlength{\abovecaptionskip}{4pt plus 3pt minus 2pt}
  \caption{Examples of ChatGPT images. The upper part is the prompt section, and the bottom is the response section in both images.}
  \label{fig:chatgpt_images} 
\end{figure}

In RQ3, we aim to understand how people interact with generative AI tools, especially ChatGPT. 
To this aim, we analyze screenshots of interactions with ChatGPT, which many users share on Twitter~\cite{TheBrill64:online}. 
These images typically feature a single-color background separating the prompt and response sections, with two primary variations: light mode and dark mode (Figure~\ref{fig:chatgpt_images}). 
This design allows us to analyze the images by simple rule-based methods.  \looseness=-1

First, we identify whether or not these images are screenshots of the ChatGPT interface. 
In essence, we consider the image a ChatGPT screenshot if two of the top three RGB colors of the image correspond to the range of RGBs of the prompt and response sections (the detail of RGBs are omitted for the reason of space).
We use the Python package Pillow~\cite{clark2015pillow} for RGB extraction. 
This simple approach achieves an F1 score of 0.93 when tested on 300 randomly sampled and manually annotated images. 
The primary cause of misclassifications arises when ChatGPT generates code--the edges of the code section share the same RGB values as the prompt section.
We solve this problem on a rule basis in text extraction since the text on the edge of codes is the only name of the language and ``Copy code.'' \looseness=-1



Next, we extract prompts from the ChatGPT images. 
We isolate the prompt section and then apply Optical Character Recognition (OCR) technique to the image. 
For OCR, we use the Python package pytesseract~\cite{hoffstaetter2021pytesseract}. 
Since pytesseract frequently produces errors when processing dark-mode images, we address this issue by inverting the RGB values for night-mode images. 
As a result, we get 93,623 prompts from ChatGPT images. \looseness=-1

\section{RQ1: Which Occupation Mention Generative AI?}\label{RQ1}

To answer RQ1, we count the users of each occupation who mention each generative AI tool.
We employ the count of users, not tweets, to mitigate the influence of vocal users.
As these counts are influenced by the inherent distribution of occupations on Twitter, we normalize this by employing the distribution obtained from $T_{rand}$ ($\S$\ref{randomsample}). 
This enables us to compute the \textit{relative presence} of each occupation that reflects the interest in generative AI more accurately.
The relative presence can be expressed as follows:
\begin{equation}
    Relative Presence_{i} = \frac{|U(T_{genAI}^i)| /|U(T_{genAI})|}{|U(T_{rand}^i)| / |U(T_{rand})|},
\end{equation}
where $|U(T_{genAI})|$ and $|U(T_{rand})|$ are the numbers of users in $T_{genAI}$ and $T_{rand}$, $i$ is a particular occupation, $|U(T_{genAI}^i)|$ and $|U(T_{rand}^i)|$ are the numbers of users of occupation $i$ in $T_{genAI}$ and $T_{rand}$, respectively. 
This measure reveals how much each occupation's presence in discussing generative AI differs from its usual engagement on Twitter.
If the value is greater than 1, it signifies that the occupation exhibits a higher presence in tweets about generative AI than other usual topics. \looseness=-1

\begin{figure}[!htbp]
\centering
\includegraphics[width=0.9\linewidth]{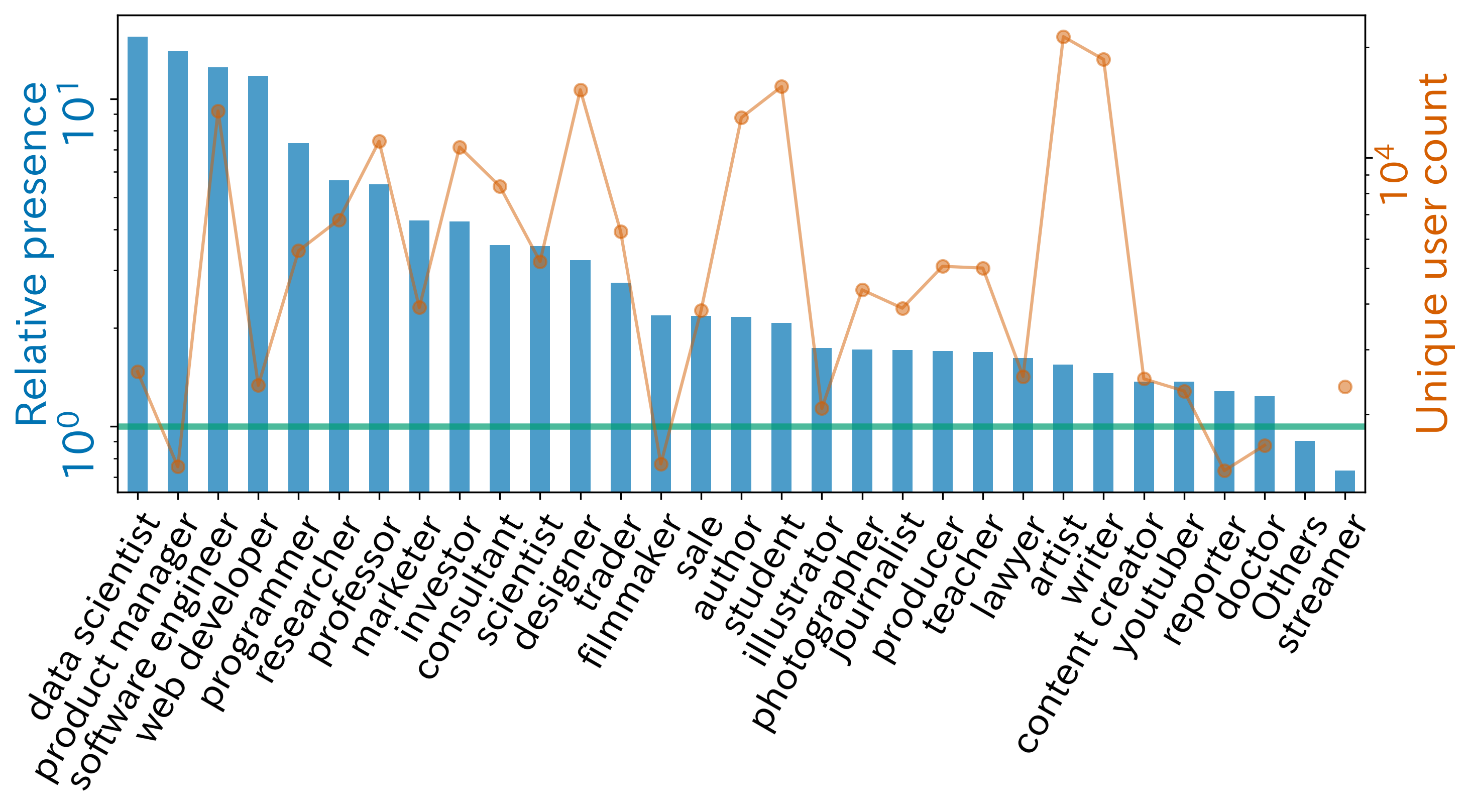}
\caption{
The relative presence and unique user count of each occupation for all generative AI tools.  
The horizontal line indicates 1 of relative presence.
The user count of Others is omitted for visibility.
}
\label{fig:bar_presence_all}
\end{figure}

Figure~\ref{fig:bar_presence_all} shows the relative presence concerning all generative AI tools by each occupation, along with the corresponding number of unique users. 
For reference, we calculate the relative presence for tweets that do not include the top 30 occupations and include them as ``Others.''
The data reveals that most occupations in this study exhibit a relative presence greater than 1, indicating their higher presence than usual.
Notably, attention to generative AI is evident not only among IT-related occupations, but also among those with less connection to IT, including lawyers, sales, traders, YouTubers, and teachers; however, only streamers show a relative presence lower than 1.
This suggests that generative AI has captured the interest of a wide range of occupations. \looseness=-1

\begin{figure}[!htbp]
\centering
\includegraphics[width=0.99\linewidth]{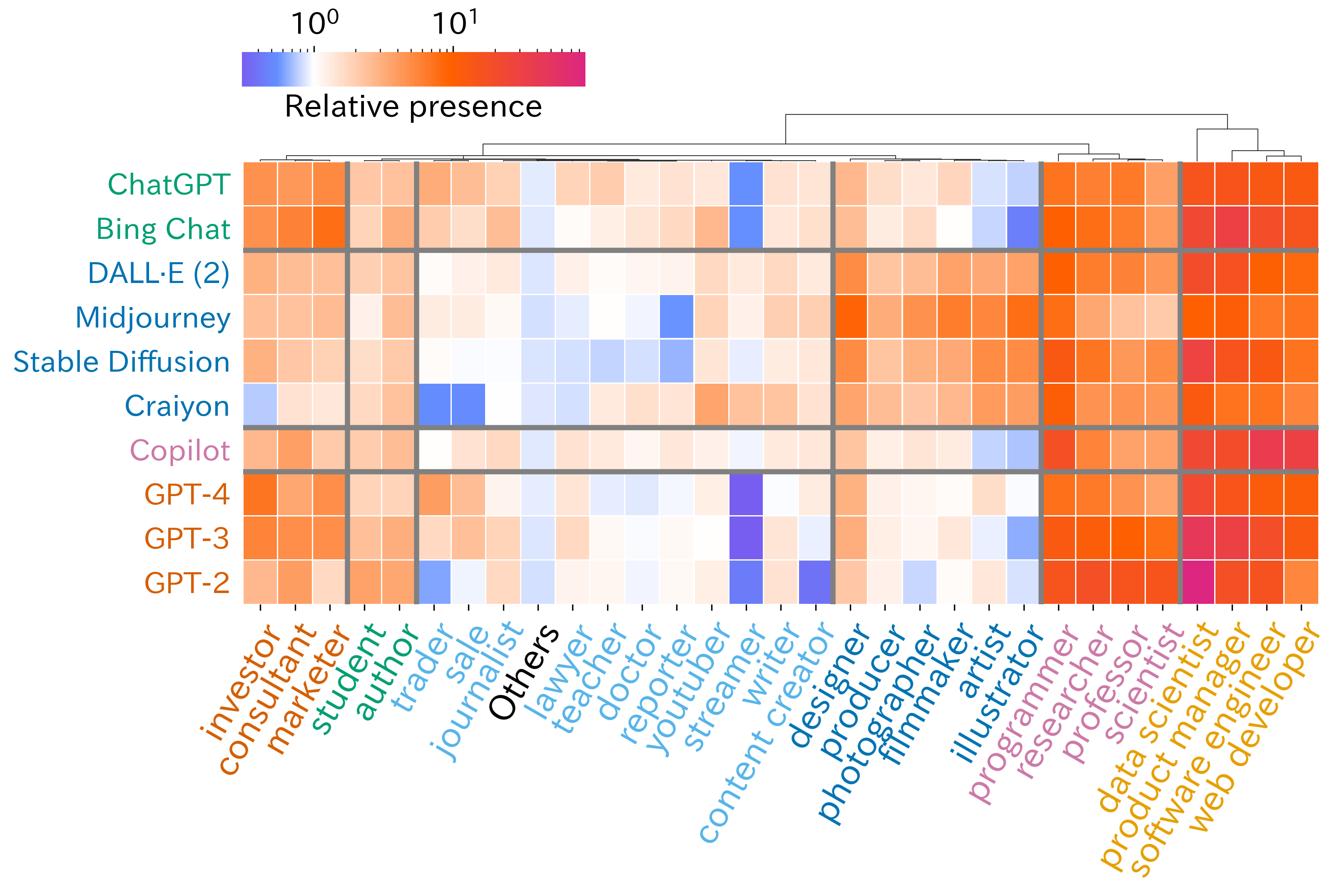}
\caption{
The relative presence of each occupation for each generative AI tool.
A thick line is drawn on the border between occupational clusters and AI categories.
}
\label{fig:heatmap_presence}
\end{figure}

Figure~\ref{fig:heatmap_presence} presents the relative presence of each generative AI tool across different occupations.
The predominance of warmer hues (values exceeding 1) indicates that most occupations discuss the variety of generative AI tools more than usual. 
We sort the occupations by Ward's hierarchical clustering method, thereby creating distinct clusters. 
IT-related occupations (e.g., data scientists), researchers, and business-related groups (e.g., marketers) express a broad interest in generative AI, although their levels of interest vary.
Conversely, occupations associated with visual content (e.g., designer and photographer) display a heightened interest in image-generative AI tools. 
For the remaining occupations, there is largely an increase in interest in various generative AI tools. 
Overall, the correspondence between clusters of occupations and the generated AI tools seems reasonable.


\section{RQ2: What are the Sentiments of Different Occupations Toward Generative AI?}\label{RQ2}
\subsection{Sentiments of Each Occupation}\label{sec_sentiments}
To investigate the perception of each occupation toward generative AI, we analyze their sentiment about generative AI tools. 
We employ a RoBERTa-based model pre-trained on Twitter data~\cite{barbieri-etal-2020-tweeteval} to classify tweets into three classes: Positive, Neutral, and Negative.
We manually validated this model with 99 tweets, 33 tweets for each class, resulting in an F1 score of 0.768, which is comparable to the reported performance of the model~\cite{barbieri-etal-2020-tweeteval}.
Since this model can output the probabilities of each model, to facilitate comparison between tweets, we aggregate the probabilities into a single value per tweet, namely sentiment score, by subtracting the Negative probability from the Positive probability (eventually, the score is scaled from -1 to 1). 
This value approaches 0 if the Positive and Negative values have similar values or if the Neutral probability gets greater. \looseness=-1

We assign this sentiment score for every tweet in $T_{genAI}$ and then aggregate them regarding each generative AI tool by occupation. 
To reduce the influence of vocal users, we first average the scores over the same users and then average them over all users. 
To account for the inherent bias in sentiment by occupation, we also compute the sentiments of randomly sampled tweets, $T_{rand}$, by occupation. 
\looseness=-1

\begin{figure*}[!ht]
\centering
  \begin{subfigure}[t]{.24\linewidth}
    \centering
    \includegraphics[width=\linewidth]{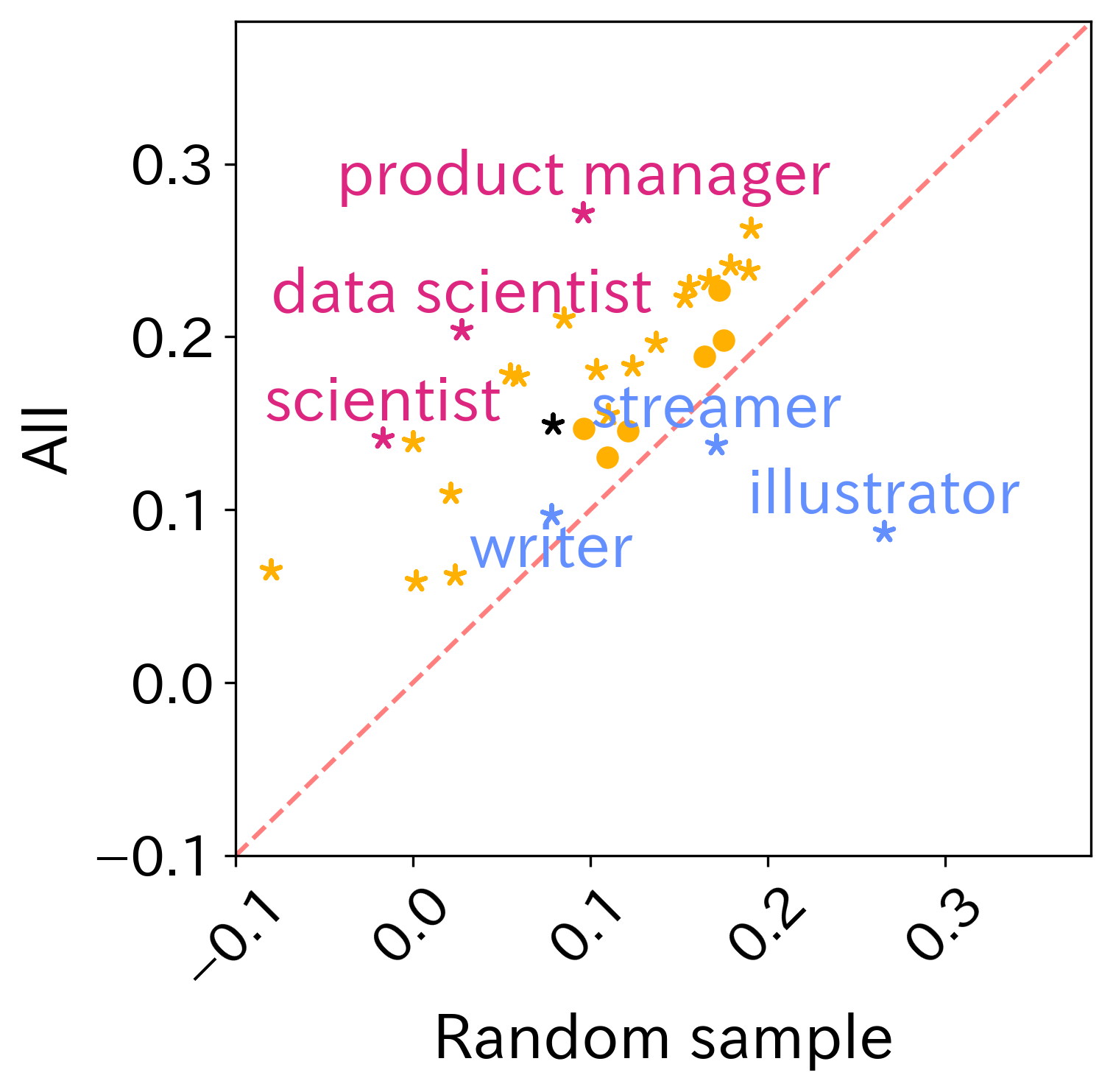}
    \caption{All generative AI tools.}
    \label{fig:2a_sb}
  \end{subfigure}
  \begin{subfigure}[t]{.24\linewidth}
    \centering
    \includegraphics[width=\linewidth]{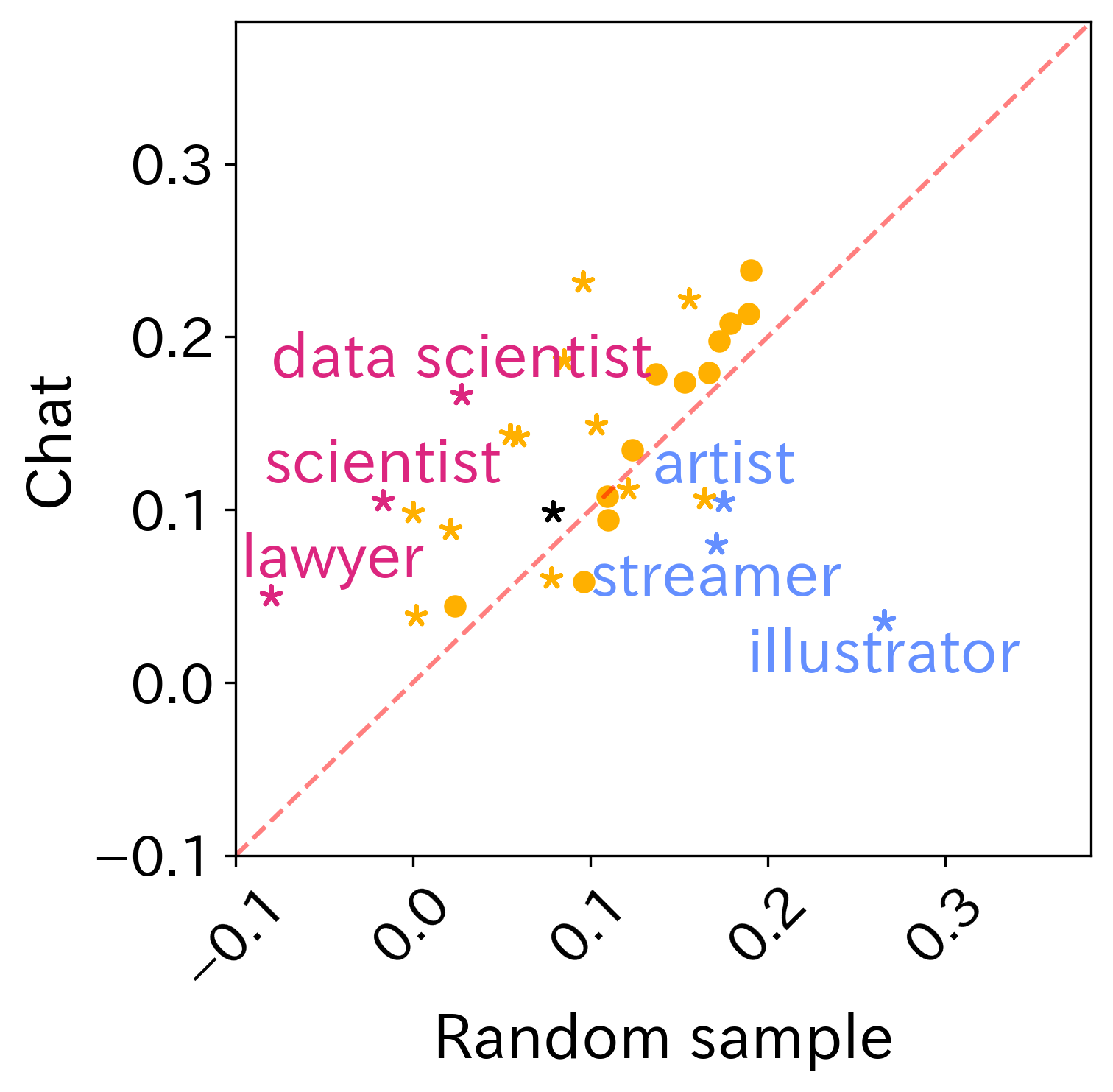}
    \caption{Chat.}
    \label{fig:2b_sb}
  \end{subfigure}
  \begin{subfigure}[t]{.24\linewidth}
    \centering
    \includegraphics[width=\linewidth]{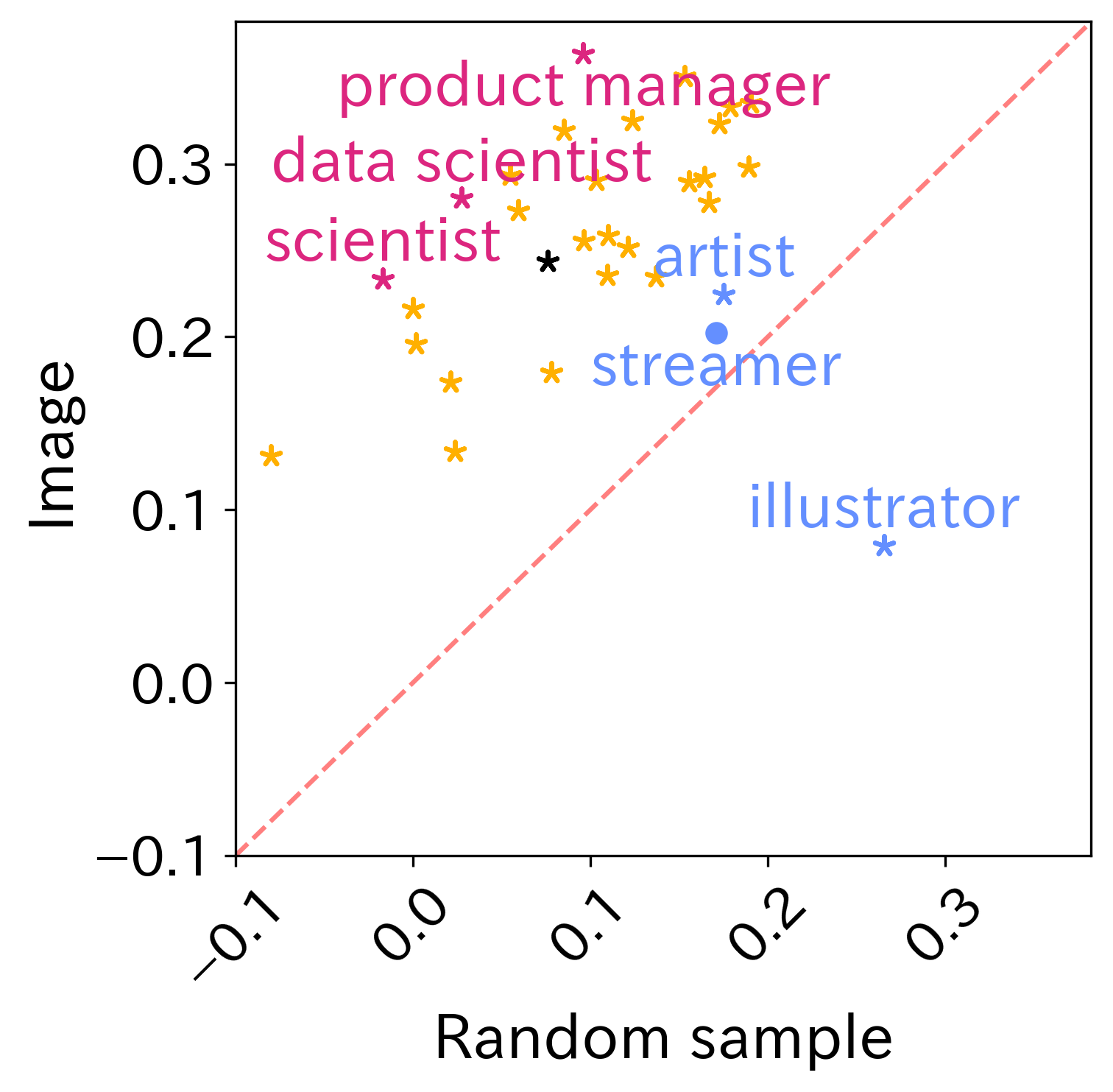}
    \caption{Image.}
    \label{fig:2c_sb}
  \end{subfigure}
  \begin{subfigure}[t]{.24\linewidth}
    \centering
    \includegraphics[width=\linewidth]{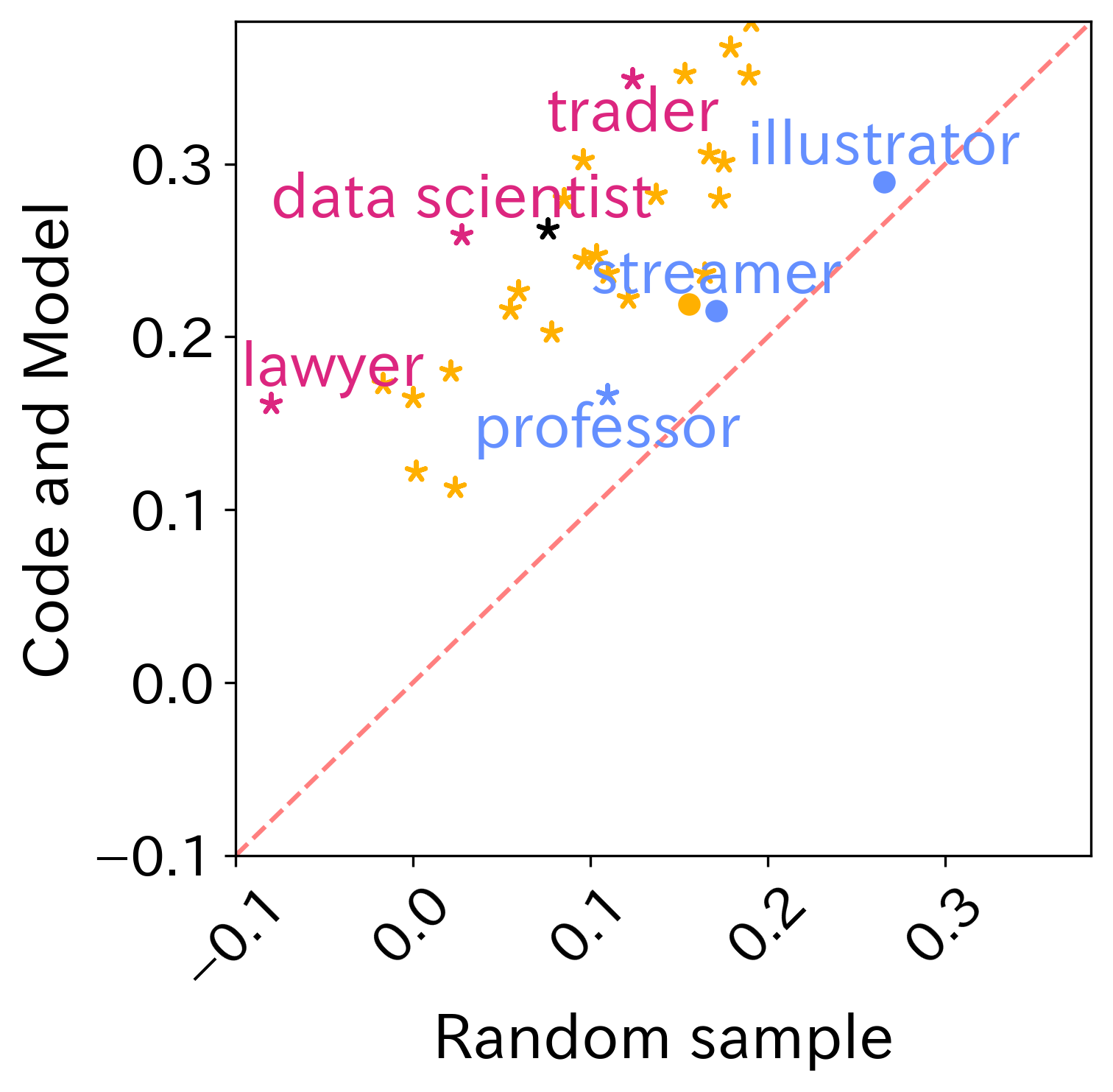}
    \caption{Code and Model.}
    \label{fig:2d_sb}
  \end{subfigure}
  \setlength{\abovecaptionskip}{4pt plus 3pt minus 2pt}
  \caption{Comparison of the mean sentiment scores toward generative AI tools by occupation and their usual sentiment. Each plot indicates occupation, with only a black mark for all tweets.
  The occupations with significant differences between the two scores are marked with a star ($p<0.05$ by the Mann-Whitney U test).}
  \label{fig:senti_scatter} 
\end{figure*}

Figure~\ref{fig:senti_scatter} displays the comparison of the sentiment of each occupation with their usual sentiments for all generative AI tools as well as for the Chat, Image, and Code/Model categories ($\S$\ref{selection}).
The $y$-axis represents the sentiment in $T_{genAI}$, while the $x$-axis shows the sentiment of tweets in $T_{rand}$. 
For clarity, we label a few occupations that have the most pronounced differences in sentiment. 
Specifically, after standardizing each value to the [0,1] scale, we divide the sentiment score in $T_{genAI}$ by that in $T_{rand}$ (called the sentiment ratio) and highlight the top and bottom three values each. 
The red dotted line in the figure indicates the $y=x$ line. 
We also plot the black marks for all tweets for reference. \looseness=-1 

The results reveal that most occupations are located above the red line, suggesting that they have positive sentiments about generative AI.
By category, Image and Code/Model exhibit highly positive sentiment compared to Chat. 
Some occupations are skewed toward negative sentiments for Image and Chat, while all occupations are more positive than usual for Code/Model.
The occupations expressing particularly positive sentiment include product managers, data scientists, scientists, traders, and lawyers. 
Lawyers, interestingly, have a more positive sentiment, despite having a relatively lower sentiment usually.
On the other hand, occupations with comparatively negative sentiments include illustrators, streamers, artists, and writers.   
In particular, illustrators are the most negative by a significant margin. \looseness=-1

To explore the factors that contribute to the sentiment of various occupations, we examine their posts by using a topic model.
There are two primary options for topic models: probabilistic topic models, such as well-known LDA, and transformer-based methods, such as BERTopic~\cite{grootendorst2022bertopic}, a combination of transformer-based text embedding and density-based clustering HDBSCAN~\cite{McInnes2017}, which can identify cohesive groups of texts in the embedding space.
The former is considered effective for classifying broad topics, while the latter is for extracting more detailed topics~\cite{ebeling2022analysis}. 
Our preliminary analysis shows that the latter, BERTopic, yields more interpretable results.
Since BERTopic yields the various number of clusters depending on the datasets and the parameters (we set it as default), we focus on the largest 5 topics. \looseness=-1

\begin{figure}[!htbp]
\centering
\includegraphics[width=0.85\linewidth]{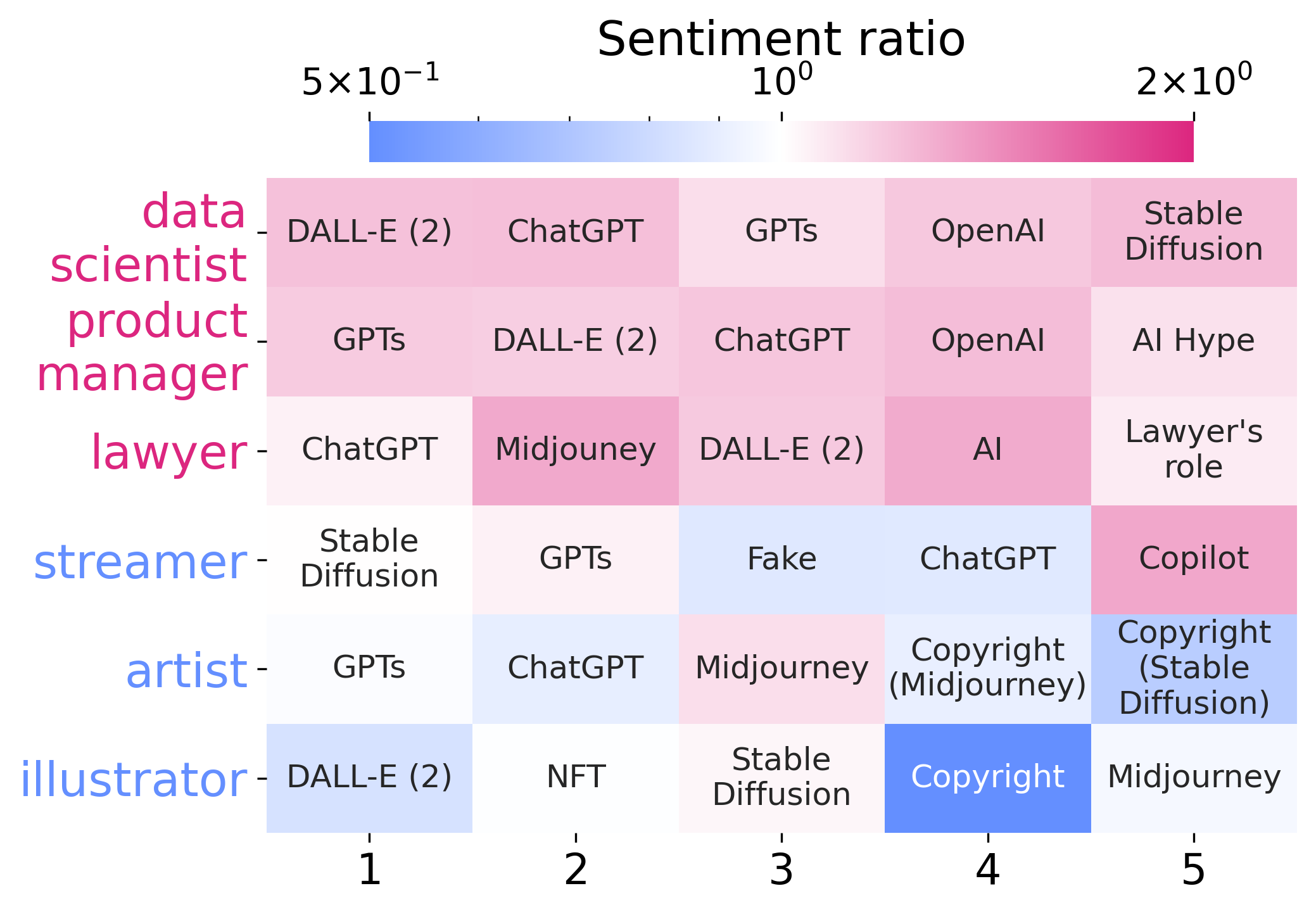}
\caption{
Top 5 topics by occupations and their sentiment. 
}
\label{fig:topic_sentiment}
\end{figure}

Figure~\ref{fig:topic_sentiment} displays the top 5 topics and their sentiment by occupations that have high or low sentiment ratios. 
The topic labels are manually assigned by the authors based on the outputs of BERTopic. 
We select the most pronounced occupations in the difference in sentiment ratio.
Occupations with positive sentiments generally praise generative AI tools or admire the progress of AI. 
As a more specific topic, lawyers discuss their roles in relation to generative AI~\cite{Newrepor32:online}.
Regarding negative sentiment, illustrators and artists highlight concerns related to copyright issues, expressing notably negative sentiments. 
An example of tweets is as follows:
\begin{quote}
``Stable Diffusion uses datasets based on art theft. Don't pretend you're doing the right thing here, CSP. You know full well this is a shitty, unethical move.''
\end{quote} 
\looseness=-1

\subsection{Comparison to AI Exposure Score}

AI's impact on occupation is a crucial topic of the relationship between humans and AI. 
The fear of emergent technology displacing occupations has been a historical concern~\cite{cave2019scary}. 
As exemplified by the Luddite movement~\cite{hobsbawm1952machine}, people may express fear and hostility toward automation technologies that threaten their jobs, which is perhaps related to the negative sentiments of each occupation. \looseness=-1

Here, we use AI exposure scores, which are the proxy of the potential impact of AI on occupations, and compare them with the sentiment scores obtained in this study. 
We utilize two types of AI exposure scores: one by \citet{webb2019impact}, who considered the overlap between job task descriptions and patent texts, namely \emph{text-based} AI exposure score, and another by \citet{felten2021occupational}, who conducted a crowdsourced survey on the relationship between AI and occupational tasks, namely \emph{survey-based} AI exposure score. 
If these scores are higher for an occupation, it is more exposed to AI.
According to the original papers, the former is an indicator of ``exposure to automation'', and the latter is ``agnostic as to whether AI substitutes for.'' 
Following \cite{eloundou2023gpts}, this study does ``not distinguish between labor-augmenting or labor-displacing effects,'' and uses them as just indicators of the strength of the relationship between AI and occupation.
These scores are linked to the occupation list from the BLS, making them applicable to our research. \looseness=-1

We match our top 30 occupations with the BLS dictionary and get their text-based and survey-based AI exposure scores. 
Since our occupation list incorporates the Indeed dictionary in addition to the BLS, some occupations cannot be matched. 
Also, if more than one occupation is matched, we use their mean average (e.g., ``designer'' matches both ``Fashion Designers'' and ``Graphic Designers''). 
For the corresponding sentiment, we use the sentiment ratio, which is the sentiment score for $T_{genAI}$ adjusted by the one for $T_{rand}$ ($\S$\ref{sec_sentiments}). \looseness=-1 

\begin{figure}[!ht]
\centering
  \begin{subfigure}[t]{.44\linewidth}
    \centering
    \includegraphics[width=\linewidth]{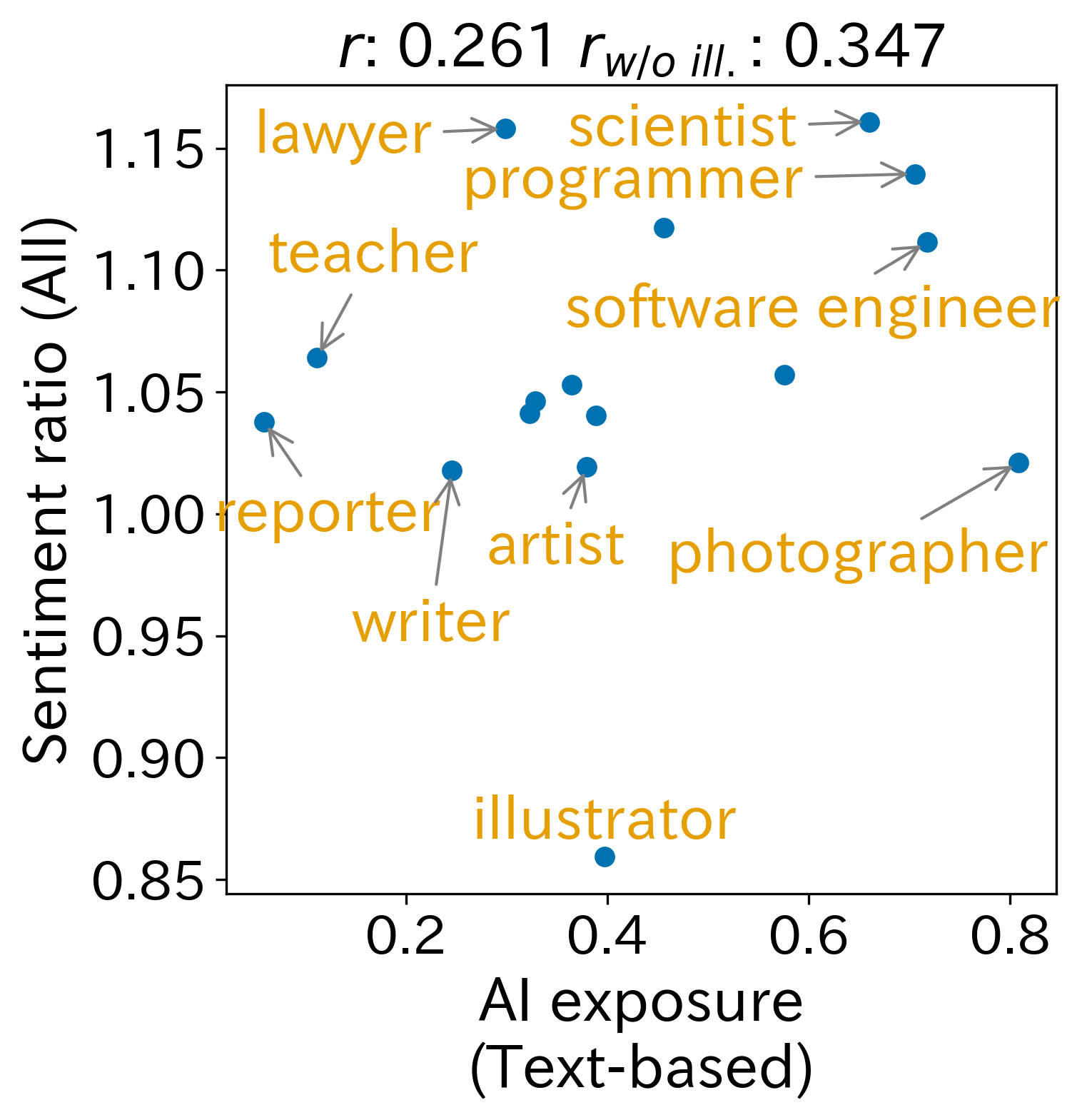}
    \caption{Text-based score}
    \label{fig:3a_sa}
  \end{subfigure}
  \begin{subfigure}[t]{.47\linewidth}
    \centering
    \includegraphics[width=\linewidth]{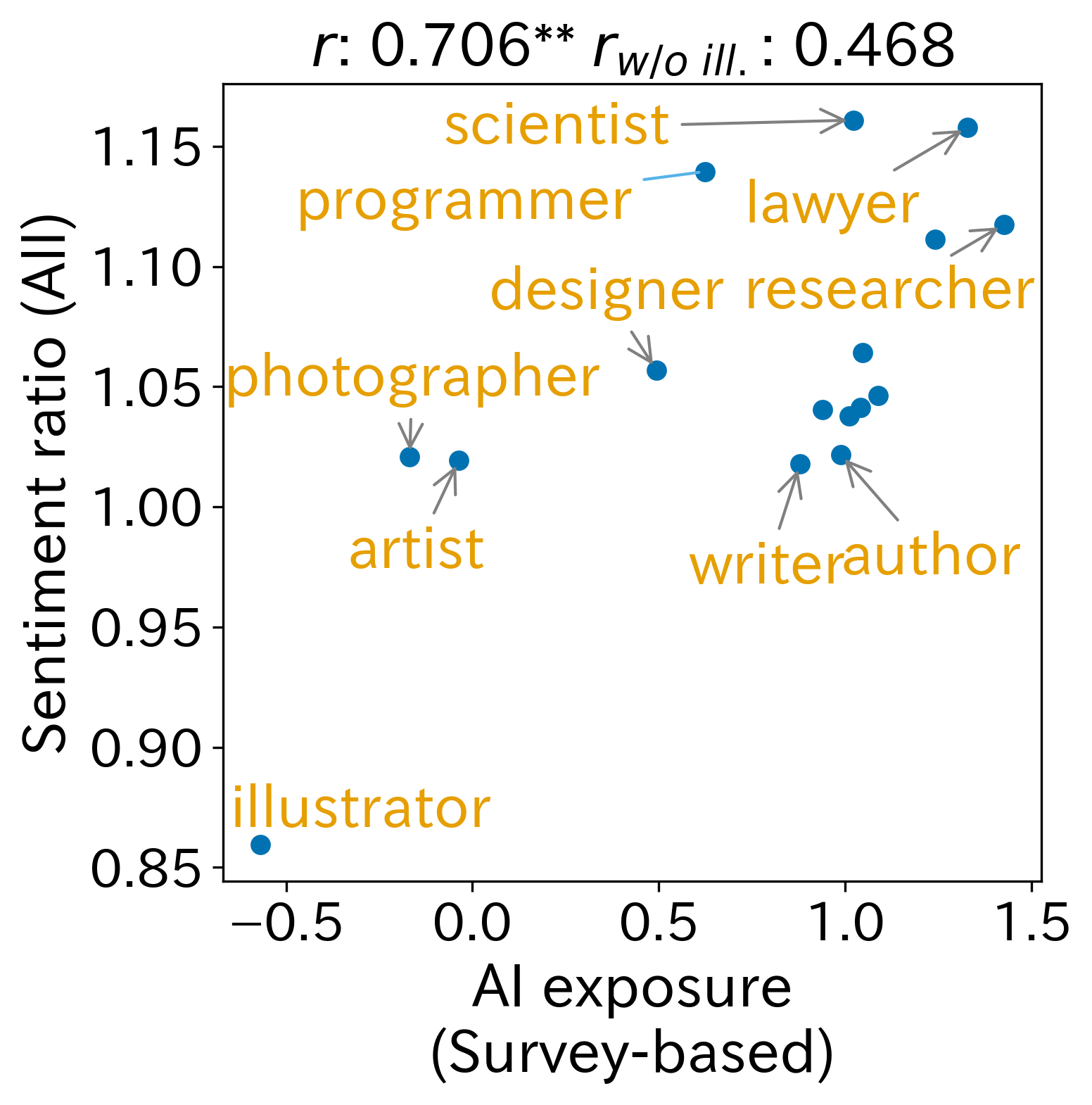}
    \caption{Survey-based score}
    \label{fig:3b_sb}
  \end{subfigure}
  \setlength{\abovecaptionskip}{4pt plus 3pt minus 2pt}
  \caption{AI exposure score and sentiments. 
  $r$ and $r_{w/o\ ill.}$ indicate the Pearson correlation score and the score without illustrators.
  ** indicates $p<0.01$.
  }
  \label{fig:senti_ai_score} 
\end{figure}

Figure~\ref{fig:senti_ai_score} shows the results.
Note that not all occupations are annotated for the reason of space.
We observe a positive correlation (albeit not always significant). 
For the text-based AI exposure score, the Pearson correlation is 0.262 ($p=0.326$), and without the illustrator, a seeming outlier, it is 0.344 ($p=0.209$). 
For the survey-based score, the Pearson correlation is 0.706 ($p=0.002$), and without the illustrator, it is 0.468 ($p=0.079$). \looseness=-1

In addition, \citet{felten2023occupational} recently created survey-based AI exposure scores specifically focused on generative AI for image and language; thus, we analyze the relationships of these image and language AI exposure scores with the sentiment ratios for the Image and Chat generative AI tools, respectively. 
The result confirms also a positive correlation. 
For image-generative AI, the Pearson correlation is 0.315 ($p=0.235$), and without the illustrator, it is 0.404 ($p=0.135$). 
For language-generative AI, the Pearson correlation is 0.731 ($p=0.001$), and without the illustrator, it is 0.557 ($p=0.031$). 
We observe a notably high correlation for language-generative AI, potentially due to people easily associating it with a labor-augmenting feature.\looseness=-1

%


In summary, users of the occupations in our dataset largely have positive sentiments toward generative AI. 
Moreover, those who have higher exposure to AI tend to have more positive sentiments. \looseness=-1

\section{RQ3: How Do People Interact with Generative AI?}
Understanding how people interact with generative AI tools is crucial for gaining insights into how they accept them. 
To this end, we focus on tweets about ChatGPT 
because it has a significant influence on global interests in AI ~\cite{WhatisCh30:online}, and also the unique habit of people posting screenshots of their usage of ChatGPT~\cite{TheBrill64:online} provides an invaluable resource for our investigation. \looseness=-1

We classify ChatGPT prompts ($\S$\ref{image_extraction}) into topics using BERTopic. 
Similar to RQ2, our preliminary analysis finds that the topics of prompts are widely varying; thus, HDBSCAN in BERTopic is beneficial as it allows us to gather groups of documents that form semantically dense clusters while treating documents in sparse positions as outliers. 
The top 10 topics are presented in Table~\ref{fig:topics_chatgpt}.
Examples include 1) work assistance such as translation and tweet writing, 2) casual and entertaining applications like rap, games, poetry, and ASCII art, and 3) more serious topics such as COVID-19 and AI models. \looseness=-1

\begin{table*}[!ht]
\centering
\scalebox{0.85}{
\begin{tabular}{lcp{7cm}p{8cm}}
\hline
Topic                    & Count & Topic words                                                                                           & Representative   text                                                              \\ \hline
Covid-19 vaccination     & 650   & vaccine, covid, covid19, virus, pandemic           & Are the covid vaccines safe?                                                       \\
Rap music composition    & 637   & rap, lyric, song, verse, eminem                                  & Write me a rap song in the style   of eminem                                       \\
Multilingual translation & 583   & translate, spanish, english, japanese, translation & Translate that into English,   please.                                             \\
Role-playing games       & 547   & game, adventure, rpg, play, zork                               & do you want to play a game with me                                                 \\
Rhymed poetry            & 445   & rhyme, poem, rhyming, poetry, twofer                              & write a poem about Al                                                            \\
AI language model        & 408   & large, language, model, trained, assistant                & do you or any other large   language model understand what you say?                \\
Tweet writing    & 363   & tweet, chatgpt, twitter, post, thread                 & Write a Tweet about   ChatGPT                                                      \\
Personal identification  & 359   & chris, who, matt, giuseppe, sean                              & But who is exactly this $\langle$personal name$\rangle$?                                       \\
Cooking and recipes      & 352   & recipe, ingredient, dish, meal, garlic                              & using only these ingredients   find me an easy recipe for a meal. beans and bread. \\
ASCII art creation       & 347   & ascii, art, draw, drawing, face                              & do you know ascii art?                                                             \\ \hline
\end{tabular} 
}
\setlength{\abovecaptionskip}{4pt plus 3pt minus 2pt}
\caption{
Top 10 topics of prompts to ChatGPT.
We hide personal names with $\langle$personal name$\rangle$.
 }
\label{fig:topics_chatgpt}
\end{table*}

\begin{figure}[!htbp]
\centering
\includegraphics[width=0.99\linewidth]{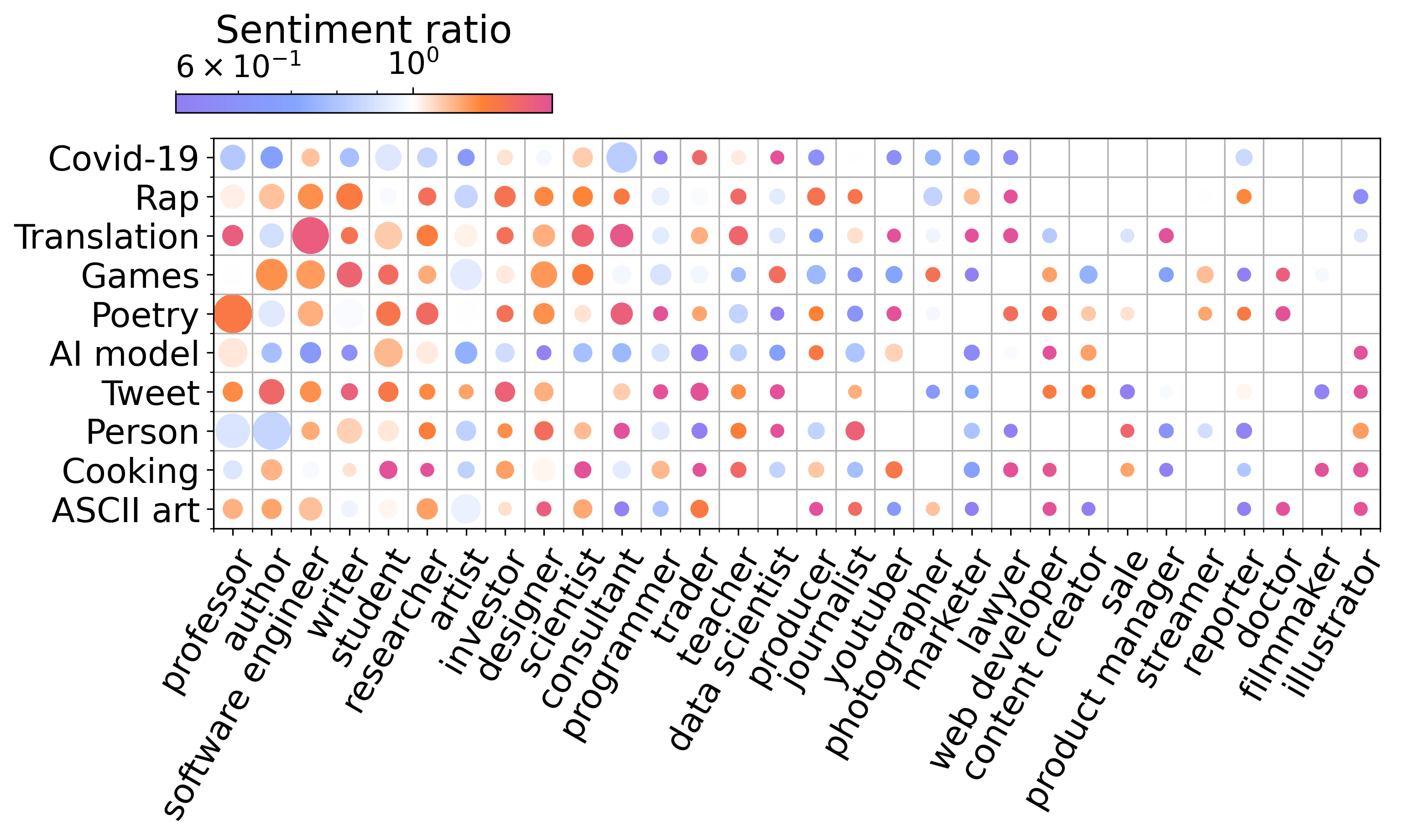}
\caption{
The relationship between occupations and the prompts to ChatGPT. 
The color indicates the sentiment ratio, and the size indicates the volume of images.
}
\label{fig:image_topics}
\end{figure}

Figure~\ref{fig:image_topics} displays the relationship between prompts and occupations. 
In this plot, the size of the circle represents the raw number of images corresponding to the occupation and prompt (minimum 1, maximum 25). 
No circle is present if the value is zero. 
The color corresponds to the sentiment ratio of the tweet texts.
Occupations are listed in the order of the volume of their ChatGPT images, with professors posting the most. \looseness=-1

We find prompts about serious topics such as COVID-19 and AI models have many circles in cooler colors, indicating a negative sentiment (the mean sentiment ratios are the lowest for these two topics, 0.871 and 0.900, respectively). 
Conversely, it is evident that many circles are of a warmer color, and most of the occupations have positive sentiments with work assistance (e.g., translation and tweets) and casual interaction with ChatGPT (e.g., poetry, rap, and ASCII art). 
This suggests that users who view ChatGPT as a workmate or a playmate tend to express positive sentiments. \looseness=-1

\section{RQ4: Has the Emergence of Generative AI Changed People's Perception of AI?}
We observe that generative AI has garnered widespread interest among the general public and is perceived positively. 
The positive reception of generative AI tools may influence people's perceptions toward AI in general. 
Consequently, we investigate the changes in perception toward AI at the collective and individual levels. \looseness=-1

\subsection{Collective Level Perception}\label{perseption_collective}
\begin{figure}[!ht]
\centering
  \begin{subfigure}[t]{.49\linewidth}
    \centering
    \includegraphics[width=\linewidth]{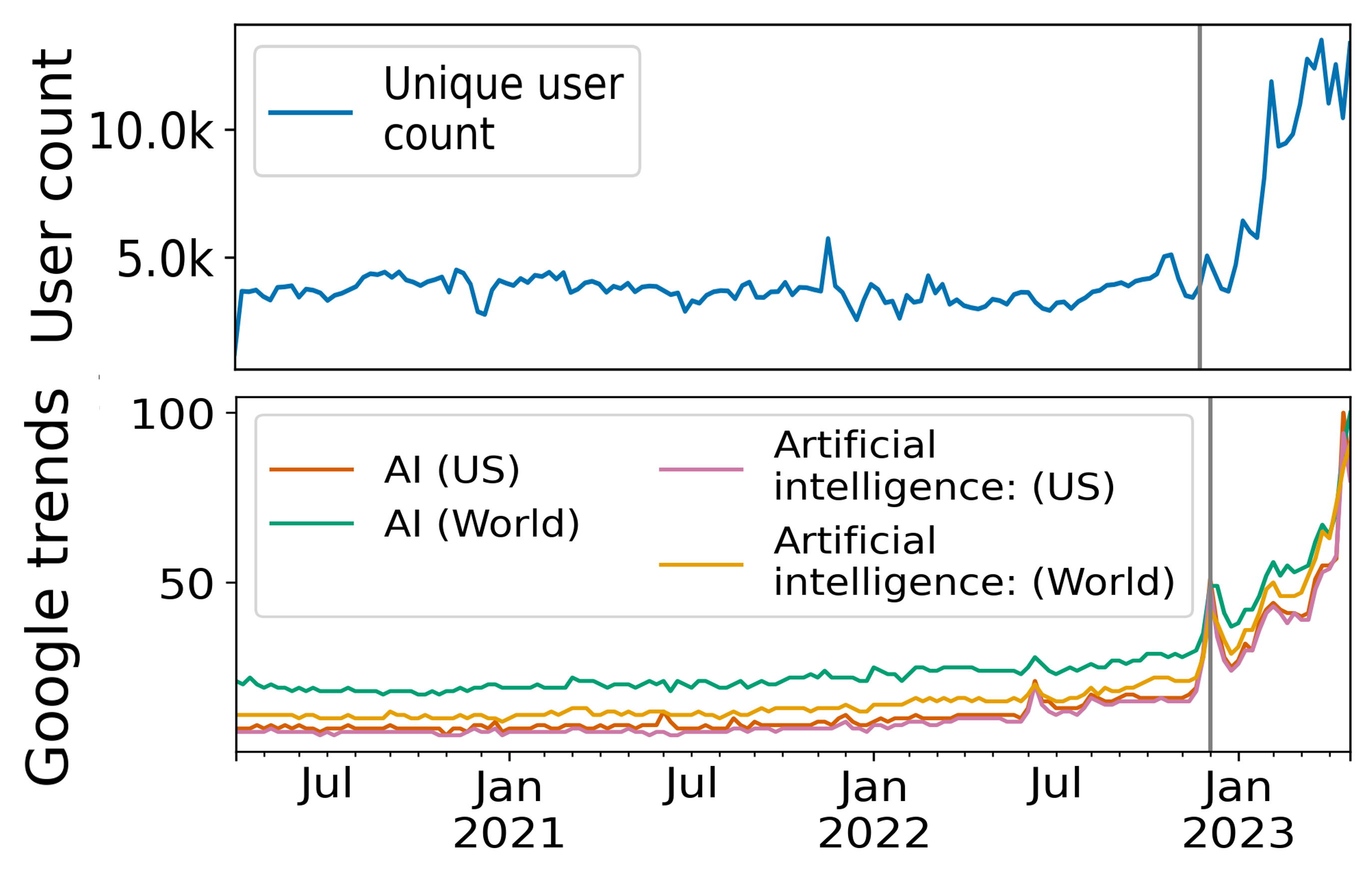}
    \caption{User count and Google trends. }
    \label{fig:5a_sa}
  \end{subfigure}
  \begin{subfigure}[t]{.46\linewidth}
    \centering
    \includegraphics[width=\linewidth]{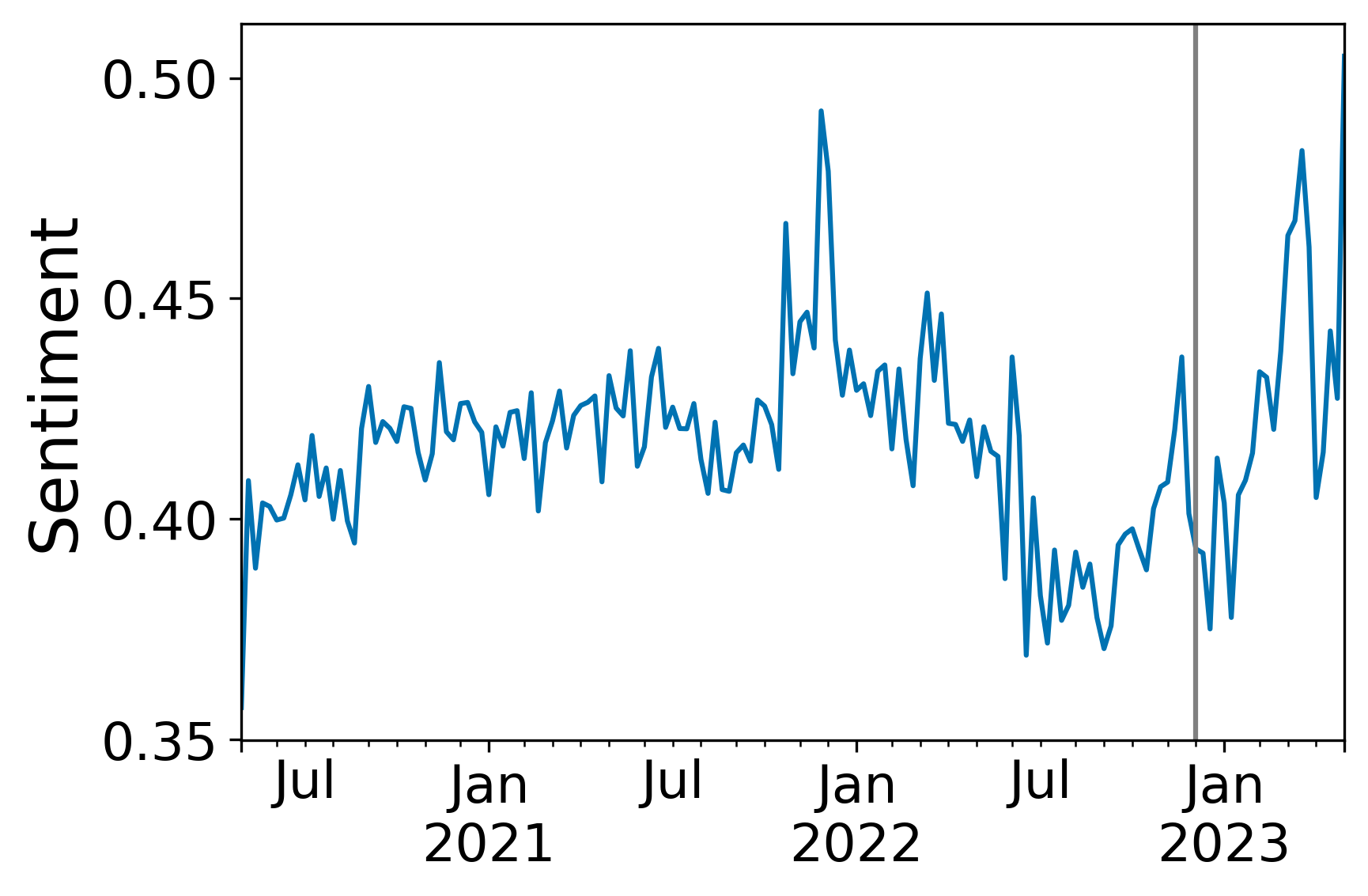}
    \caption{Sentiments.}
    \label{fig:5b_sb}
  \end{subfigure}
  \setlength{\abovecaptionskip}{4pt plus 3pt minus 2pt}
  \caption{Temporal dynamics of randomly sampled AI-related tweets and Google Trends. The vertical line indicates the release date of ChatGPT.}
  \label{fig:ai_random} 
\end{figure}

First, from our Decahose dataset, we extract tweets that contain the ``\#AI'' and ``\#ArtificialIntelligence'' (case insensitive) following~\cite{manikonda2018tweeting,fast2017long} ($\S$\ref{randomsample}). 
The upper part of Figure~\ref{fig:ai_random}(\subref{fig:5a_sa}) shows the number of unique users posting about AI in a weekly time series, which remains consistent until around December 2022, after which it increases rapidly. 
Considering that ChatGPT was released on Nov. 30, 2022, it appears that ChatGPT has prompted people to tweet more about AI.
This trend is also observed in Google Trends (bottom part of Figure~\ref{fig:ai_random}(\subref{fig:5a_sa})), in which we use the search term ``AI'' and topic ``Artificial Intelligence'' both in the US and worldwide\footnote{\url{https://newsinitiative.withgoogle.com/resources/lessons/advanced-google-trends/}}. \looseness=-1

Figure~\ref{fig:ai_random}(\subref{fig:5b_sb}) displays the average sentiment of users in the weekly time series. 
This reveals that there is no uniform trend over the long term. 
Instead, we observe a surge in sentiment at the end of 2021 and early 2023. 
To identify the factors behind the change in sentiment, we conduct a topic model analysis, using the Biterm Topic Model (BTM)~\cite{yan2013biterm}, an LDA-based approach that specializes in short texts.
Our preliminary analysis indicated that BTM performs better in extracting comprehensive topics than BERTopic.
The number of topics was determined to be 6, based on observing changes in perplexity within the range of [3,11]~\cite{zhao2015heuristic}.
The authors manually assign the labels of topics, namely Business, AI model, Crypto, Data science, Art/NFT, and Robotics.  \looseness=-1

\begin{figure}[!ht]
\centering
  \begin{subfigure}[t]{.48\linewidth}
    \centering
    \includegraphics[width=\linewidth]{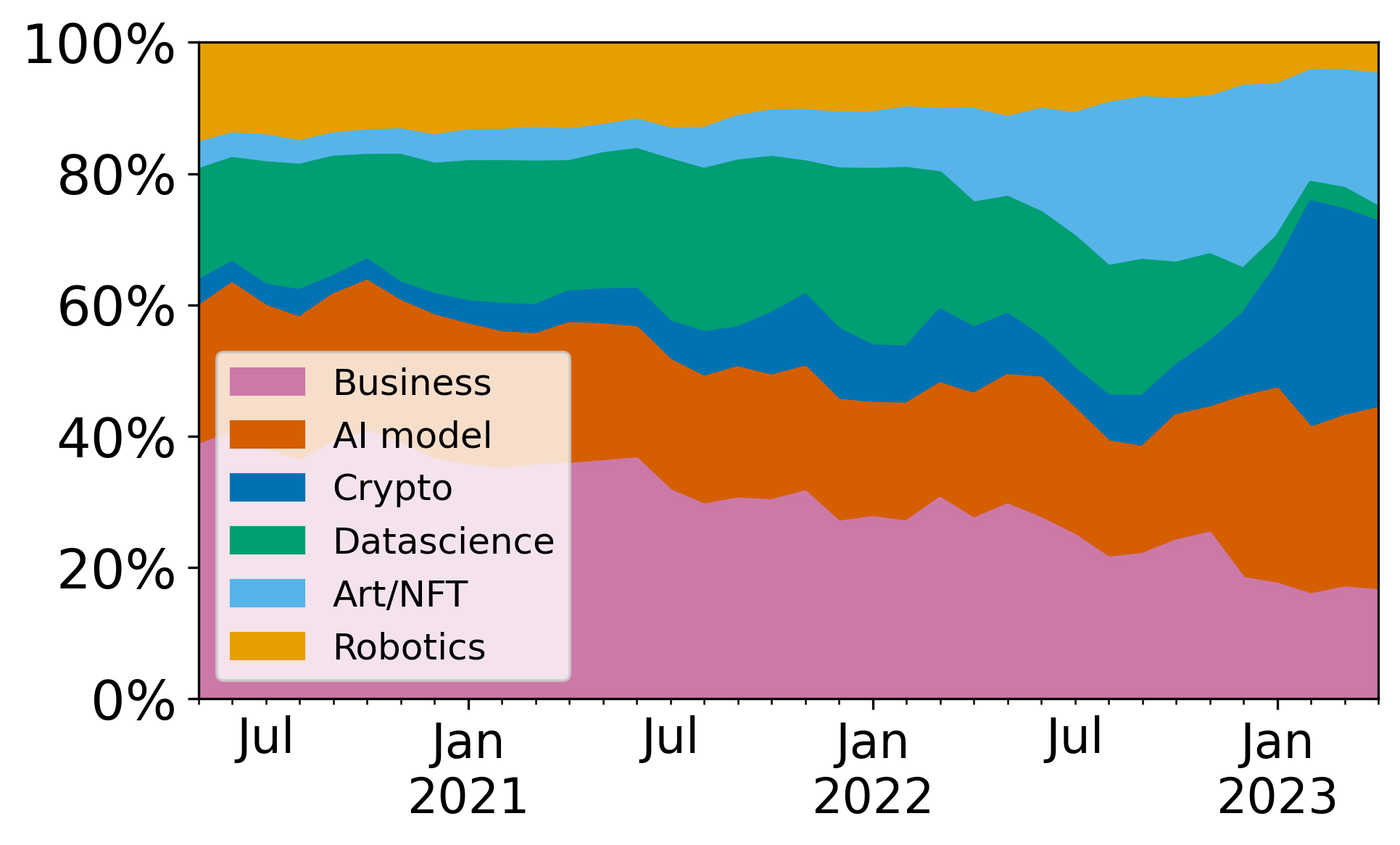}
    \caption{User count (ratio). }
    \label{fig:7a_sa}
  \end{subfigure}
  \begin{subfigure}[t]{.46\linewidth}
    \centering
    \includegraphics[width=\linewidth]{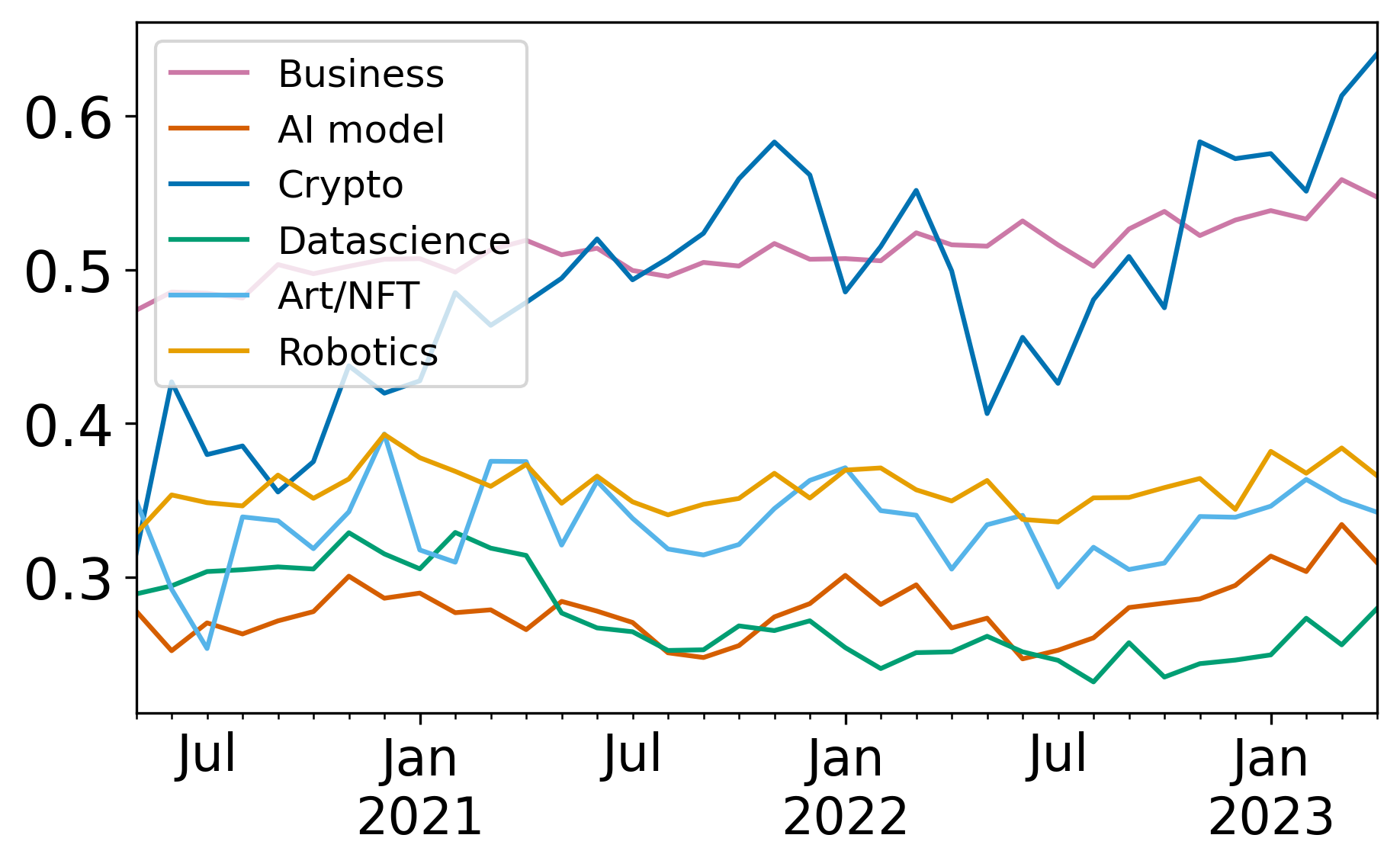}
    \caption{Sentiments.}
    \label{fig:7b_sb}
  \end{subfigure}
  \setlength{\abovecaptionskip}{4pt plus 3pt minus 2pt}
  \caption{Temporal dynamics of randomly sampled AI-related tweets by topics.}
  \label{fig:ai_random_arealine} 
\end{figure}

We assign topics to all tweets and visualize the proportions of topics 
on a monthly basis in Figure~\ref{fig:ai_random_arealine}(\subref{fig:7a_sa}).
We observe the sharpest increases in Art/NFT-related topics (early 2022) and Crypto-related topics (late 2022), which seemingly corresponds to the releases of Midjourney/DALLE-2 and ChatGPT. \looseness=-1

To confirm the impact of ChatGPT on AI-related discussions, we extract representative words for AI-related tweets before and after the release date of ChatGPT (Nov. 30, 2022) with log-odds ratios~\cite{monroe2008fightin}, using $T_{rand}$ as the background corpus.
Except for the names of generative AI tools, the top 5 most representative words are ``openai,'' ``memecoins,'' ``gpt,'' ``gem,'' and ``binance'' for tweets after ChatGPT, while they are ``data,'' ``technology,'' ``datascience,'' ``machine,'' and ``bigdata'' for tweets before ChatGPT.
This suggests that AI discussions were dominated by solid-technology topics before ChatGPT, but after ChatGPT, they are more discussed with crypto-related topics. \looseness=-1

As for sentiments, Figure~\ref{fig:ai_random_arealine}(\subref{fig:7b_sb}) shows that the only topic experiencing a sharp change is crypto-related. 
This change aligns with the overall shift in sentiment (Figure~\ref{fig:ai_random}(\subref{fig:5b_sb})), suggesting that the rapid change in AI sentiment is primarily driven by crypto-related topics, with no significant long-term trends or abrupt changes observed for the other topics. \looseness=-1

\subsection{Individual Level Perception}

To analyze users' perceptions of generative AI from a different perspective, we collect historical tweets from users who mentioned generative AI tools ($\S$\ref{hist_tweets}). 
We gather tweets containing ``\#AI'' and ``\#ArtificialIntelligence'' from 60 days before and after the day a user first mentioned a generative AI tool.
We infer their sentiment and track changes in their perceptions using the Regression Discontinuity Design (RDD) framework, which is actively used in social media research to capture the changes in multiple regimes of data~\citep[e.g.][]{miyazakiRDD}.
We employ a linear model: $V_d=\alpha_0+\beta_0 d+\alpha i_d+\beta i_d d+\epsilon_d$,
where $V_d$ is the user count or the user-based mean sentiment concerning AI on a given date $d$, and $i_d$ is an indicator variable, which is 1 after the generative AI tweets (i.e., $d>0$) and 0 otherwise.
We align and aggregate the tweets of all the users in this framework. \looseness=-1

\begin{figure}[!ht]
\centering
  \begin{subfigure}[t]{.48\linewidth}
    \centering
    \includegraphics[width=\linewidth]{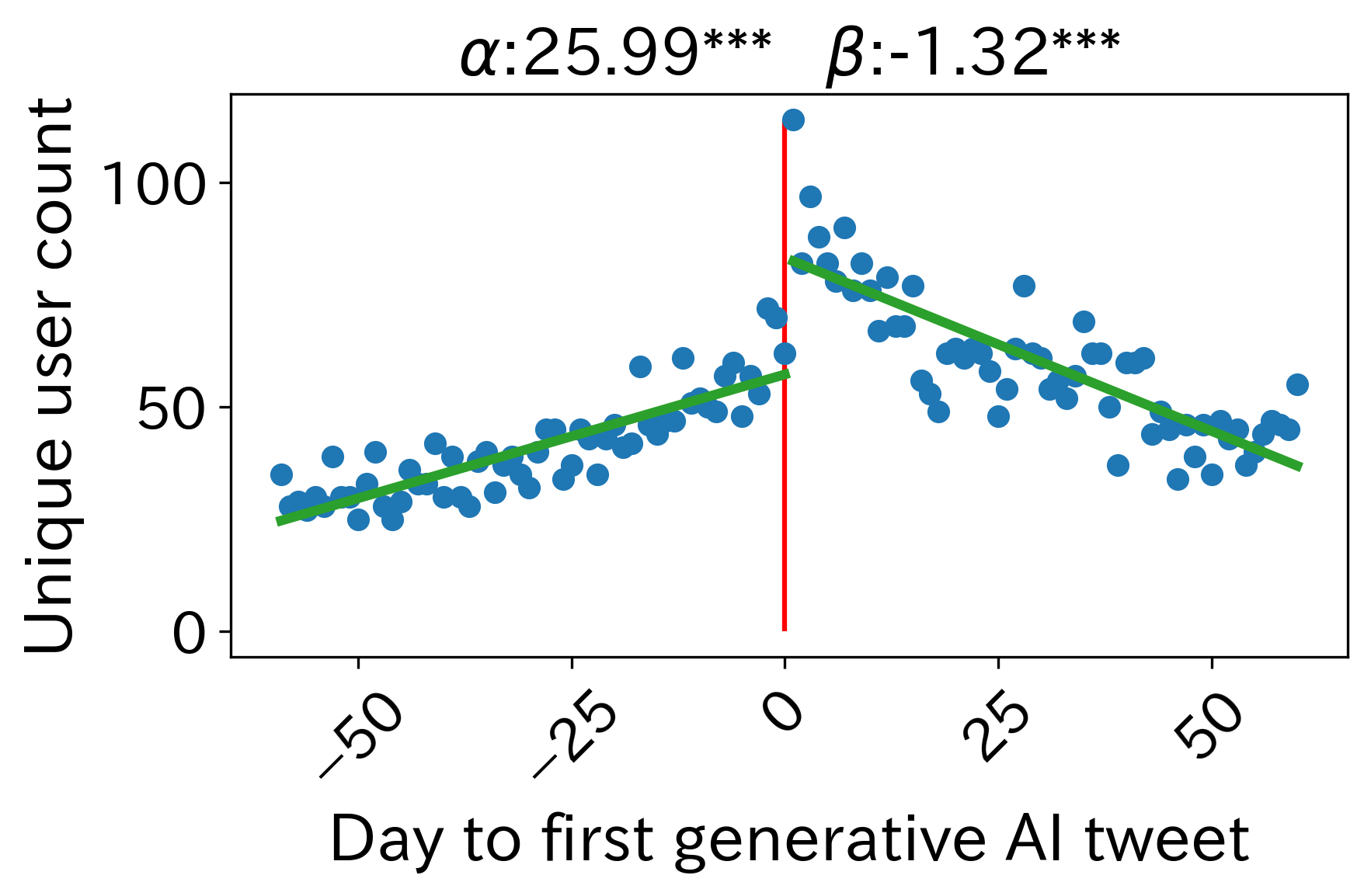}
    \caption{User count.}
    \label{fig:6a_sb}
  \end{subfigure}
  \begin{subfigure}[t]{.48\linewidth}
    \centering
    \includegraphics[width=\linewidth]{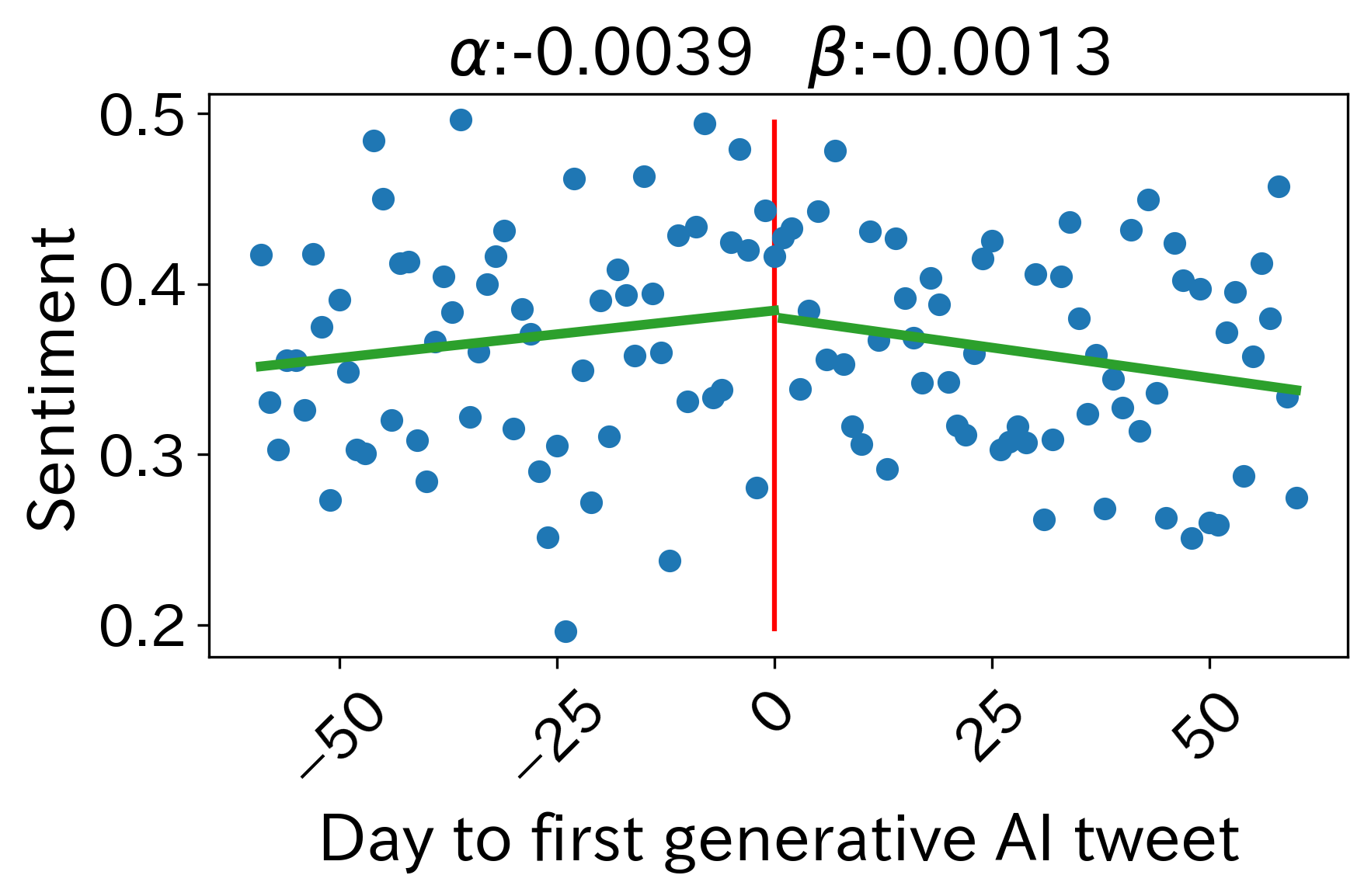}
    \caption{Sentiment.}
    \label{fig:6b_sb}
  \end{subfigure}
  \setlength{\abovecaptionskip}{4pt plus 3pt minus 2pt}
  \caption{The changes about users' AI-related tweets before and after their first tweet about generative AI.}
  \label{fig:rdd} 
\end{figure}

Figure~\ref{fig:rdd}(\subref{fig:6a_sb}) shows a jump ($\alpha$) in the user counts mentioning AI after their tweets about the generative AI tool, although there is a slight reduction ($\beta$) over the 60 days after that. 
Conversely, we could not find any changes in sentiment (Figure~\ref{fig:rdd}(\subref{fig:6b_sb})). \looseness=-1

When we applied RDD to tweets of each generated AI tool, only Midjourney showed a significant result in terms of sentiment ($\alpha$: 0.109 with $p=0.006$). 
Conducting the same analysis for each occupation, we find that only three occupations show significant jumps: illustrator ($\alpha$: 0.422 with $p=0.029$), content creator ($\alpha$: 0.191 with $p=0.042$), and data scientist ($0.164$ with $p=0.032$). \looseness=-1

The results obtained for Midjourney and illustrators are surprising, considering the copyright issue discussed in $\S$\ref{RQ2}. 
This could be attributed to the low base sentiment of illustrators toward AI (0.179, while the average of the other occupations is 0.356 with a standard deviation of 0.128). 
Additionally, $\alpha$ in RDD just indicates the reactions immediately after the first tweet about generative AI tools, suggesting that sentiment may have initially got positive due to admiration for its performance, but got negative after a while due to the emergence of copyright issues. \looseness=-1

\section{Discussion and Conclusion}
\subsection{Main Findings}

Our investigation finds that many occupations, including non-IT-related ones, mention Generative AI more than other usual topics (RQ1). 
Also, the public interest in AI dramatically increased after the release of ChatGPT (RQ4).
In fact, AI has been relatively unfamiliar to the general public~\cite{HowAmeri78:online,cave2019scary}; however, our results show that the emergence of generative AI, particularly ChatGPT, has played a significant role in bridging the gap between AI and the general public.  \looseness=-1

The general public sentiment toward generative AI is mostly positive (RQ2), which seems to conflict with some media outlets saying people are terrified~\cite{PeopleFe45:online}.
The trend of positive sentiment toward AI is consistent with other existing studies~\cite{manikonda2018tweeting,fast2017long}; moreover, the comparison of the sentiments with ``usual tweets'' in our study further reinforces this finding. \looseness=-1

Surprisingly, occupations with higher exposure to AI exhibited more positive sentiments toward generative AI (RQ2). 
This may suggest that individuals in these occupations might be more adaptable to new technology, and their open attitude toward AI could overcome their concerns about job displacement. 
Conversely, occupations with lower exposure to AI display more negative sentiment, possibly due to their lack of preparedness for the rapid emergence of disruptive technology.
It is a well-known human trait to fear unfamiliar objects~\cite{cao2011fear}, while familiarity is thought to alleviate unnecessary fear. 
For example, the previous survey reported that the more people know about self-driving cars or related services, the more likely they are to be supportive of autonomous vehicle~\cite{horowitz2021influences} and to believe that such technology is a good idea~\cite{HowAmeri78:online}. 
In essence, if the positive perception of new technology is influenced by exposure to it, increasing exposure to new technology for the general population could help promote quicker adaptation and acceptance in the future. 
Indeed, the improvement in illustrators' sentiment toward AI in general after the first mention of generative AI may be indicative (RQ4).
This underscores the importance of familiarizing people with AI and other emerging technologies to foster a smoother transition and integration into society. \looseness=-1

On the other hand, illustrators generally have negative sentiments toward generative AI, with the primary concern revolving around the use of their artwork for training AI without consent (RQ2). 
The concerns raised by illustrators highlight the ethical implications surrounding AI technology~\cite{Japanese39:online}. 
While new technologies generally garner positive sentiments, a few topics, such as this ethical issue, can have exceptionally negative sentiments; thus, it is essential for companies and governments to address these specific concerns without overlooking them to ensure the responsible development of AI technologies.  \looseness=-1

The use of ChatGPT is often in the topics about work assistance or, interestingly, casual context and is associated with more positive sentiments (RQ3). 
Indeed, previous studies in the field of human-robot interaction suggest that active engagement in playful activities with robots can foster closer human-robot relationships~\cite{franccois2009long}. 
The finding in our study may carry implications for the evolution of human-AI relationships. 
For instance, promoting firsthand experiences with AI, such as allowing individuals to interact directly with AI systems, could be an effective strategy for enhancing the IT/AI literacy of the general population. \looseness=-1

Public interest in AI has remarkably increased, associated with a pronounced surge in crypto-related topics (RQ4). 
The post-ChatGPT period is the most prominent time for AI awareness over the past three years, underscoring the transformative impact of ChatGPT. 
However, this does not cause a significant shift in overall sentiment toward AI, with only the sentiment toward crypto-related topics showing significant improvement. 
This seems an odd result, given that generative AI and crypto seem to have a little relationship.
While it is encouraging to see AI gaining increased attention, the fact that crypto also gets a boost at the same time is a similar phenomenon to ``hashtag hijacking''~\cite{vandam2016detecting}, a free ride on the AI boost to get more public attention, such as:
\begin{quote}
``I'm so excited about all the new stuff that's happening in \#web3 and \#crypto right now! and that's not to mention \#AI etc!!!''
\end{quote}
Whatever the case, AI may have strengthened its hype-worthy nature after ChatGPT. \looseness=-1

Interestingly, the issue of hallucinations~\cite{alkaissi2023artificial}, a topic already extensively discussed in the research community, does not emerge in our analysis. 
Only streamers notably mentioned ``Fake'' as a negative aspect (RQ2). 
This suggests that, as of the time of this study,  hallucinations may not be a dominant concern in public discussions around generative AI. 
However, as generative AI becomes more widely used over time, it is plausible that this issue may gain more prominence in public discourse. \looseness=-1

\subsection{Limitations and Future Works}


\noindent \textbf{Detection and selection of occupations.}
The detection of users' occupations using Twitter profiles may not be accurate due to potential self-disclosure and falsehood~\cite{sloan2015tweets}. 
Moreover, our analysis focuses solely on the major 30 occupations, not encompassing all professions, which may be affected by inherent bias on Twitter. 
Future research should strive for a more comprehensive study, possibly leveraging external data sources, that mitigates these biases.\looseness=-1

\noindent \textbf{Twitter user bias.}
Since people on Twitter are said to be younger than the average~\cite{HowTwitt88:online}, they may have a bias with heightened information sensitivity and more forward-thinking attitudes toward new technologies.
Yet, even with this potential skew, the comparisons between occupations and the topic analysis provide meaningful insights, although mitigating these biases remains an essential task for future research. \looseness=-1

\noindent \textbf{Further comparisons of generative AI tools.}
While our study primarily focuses on occupations, it would indeed be intriguing to delve deeper into the comparison of different generative AI, particularly given the relatively negative sentiment observed towards ``Chat'' generative AI tools.

\noindent \textbf{Other types of user attributes.}
In addition to occupations, there may be other relevant factors influencing users' perception and usage of generative AI, such as age, education, number of followers, or cultural background~\cite{brossard2009religiosity}, which would provide a more nuanced understanding of how different user segments engage with generative AI in future work. \looseness=-1 

\noindent \textbf{Other languages.}
Our study focuses on English-language tweets; however, conducting an analysis of how responses to ChatGPT varied across different languages would be particularly intriguing, given the multilingual capabilities of the AI model. 
Future research could extend the analysis to include tweets in other languages and from users in different countries to explore cross-cultural differences in the perception and usage of generative AI. \looseness=-1

\section*{Broader Perspective and Ethics, and Competing Interests}   
A positive outcome of our research is that it can help policymakers to establish policies about human-AI symbiosis. 
For example, facilitating the setting in which students play with AI in education could increase AI literacy among the public. 
On the other hand, as some readers may erroneously generalize our study and mistakenly believe that only illustrators strongly oppose AI, it is important to be cautious in disregarding valid criticisms and concerns surrounding AI.
The copyright issues highlighted in this study are important and should be carefully considered.
Also, we believe there are no competing interests regarding this study.
As for privacy concerns, we did not include personal names or account names in our analysis. 
We will share only a list of tweet IDs, according to Twitter's guidelines. \looseness=-1



\bibliography{main}

\end{document}